TOPICAL REVIEW

# Tunnelling ionization of deep centres in high-frequency electric fields

S D Ganichev[1,2], I N Yassievich[2] and W Prettl[1]

[1] Institut für Experimentelle und Angewandte Physik, Universität Regensburg,
93040 Regensburg, Germany
[2] AF Ioffe Physico-Technical Institute, Russian Academy of Sciences,
194021 St Petersburg, Russia

**Abstract**
Experimental and theoretical work on the ionization of deep impurity centres in the alternating terahertz field of high-intensity far-infrared laser radiation, with photon energies tens of times lower than the impurity ionization energy, is reviewed. It is shown that impurity ionization is due to phonon-assisted tunnelling which proceeds at high electric field strengths into direct tunnelling without involving phonons. In the quasi-static regime of low frequencies the tunnelling probability is independent of frequency. Carrier emission is accomplished by defect tunnelling in configuration space and electron tunnelling through the potential well formed by the attractive force of the impurity and the externally applied electric field. The dependence of the ionization probability on the electric field strength permits one to determine defect tunnelling times, the structure of the adiabatic potentials of the defect, and the Huang–Rhys parameters of electron–phonon interaction.

Raising the frequency leads to an enhancement of the tunnelling ionization and the tunnelling probability becomes frequency dependent. The transition from the frequency-independent quasi-static limit to frequency-dependent tunnelling is determined by the tunnelling time which is, in the case of phonon-assisted tunnelling, controlled by the temperature. This transition to the high-frequency limit represents the boundary between semiclassical physics, where the radiation field has a classical amplitude, and full quantum mechanics where the radiation field is quantized and impurity ionization is caused by multiphoton processes. In both the quasi-static and the high-frequency limits, the application of an external magnetic field perpendicular to the electric field reduces the ionization probability when the cyclotron frequency becomes larger than the reciprocal tunnelling time and also shifts the boundary between the quasi-static and the frequency-dependent limits to higher frequencies.

At low intensities, ionization of charged impurities may also occur through the Poole–Frenkel effect by thermal excitation over the potential well formed by the Coulomb potential and the applied electric field. Poole–Frenkel ionization precedes the range of phonon-assisted tunnelling on the electric field scale and enhances the ionization probability at low electric field strengths. Applying far-infrared lasers as sources of a terahertz electric field, the Poole–Frenkel effect can clearly be observed, allowing one to reach a conclusion regarding the charge of deep impurities.

**Contents**

**1. Introduction**

One of the main manifestations of quantum mechanics is tunnelling. An essential feature of the tunnel effect is the absence of a time lag. The tunnelling probability remains unchanged up to very high frequencies. At higher frequencies the tunnelling process proceeds into multiphoton transitions as long as the photon energy is less than the particular activation energy. This was fully worked out theoretically for the first time by Keldysh [1]. The development of high-power terahertz lasers allowed experimental investigation of the transitions between these two limits; the result found was that the rising frequency drastically enhances the tunnelling probability. This has been observed in semiconductors where the transition from the quasi-static regime, where tunnelling is independent of frequency, to the high-frequency regime, where the enhancement of tunnelling occurs, is found to be at terahertz frequencies [2]. Tunnelling in alternating potentials is important in a variety of physical phenomena such as field emission, interband breakdown, tunnelling chemical reactions, Coulomb blockade, and the destruction of adiabatic invariance. Furthermore, tunnelling in high-frequency fields is the most effective mechanism of absorption of high-power radiation in the transparency region of dielectrics between phonon and electron excitations.

This review deals with the newly found nonlinear optical effect of deep-impurity ionization by terahertz radiation with photon energies a few tens of times lower than the impurity binding energy. This effect was first observed in the photoconductive signal of semiconductors doped with deep impurities in response to high-power terahertz radiation [3]. The measurements were carried out on impurity centres with no direct coupling of light to localized vibrational modes. The ionization mechanism has been investigated in great detail [4–6] yielding the finding that deep impurities can be ionized by tunnelling through the oscillating potential well formed by a strong terahertz electric field of far-infrared (FIR) radiation together with the attractive potential of the defect. Ionization of impurities in semiconductors by dc electric fields is well known and is, in particular, applied in deep-level transient spectroscopy (DLTS) [7]. It will be shown that in a certain limit, terahertz radiation acts like a strong dc electric field ionizing deep impurities in spite of the fact that the field is alternating and the quantum energy is much smaller than the impurity binding energy. In this limit the ionization probability does not depend on the radiation frequency. An increase of the frequency and decrease of temperature result in the ionization probability becoming dependent on frequency, which indicates a transition to the range where the quantization of the radiation field becomes significant [2]. This transition takes place at $\omega\tau = 1$ where $\omega$ is the radiation frequency and $\tau$ is the tunnelling time [8] in the sense of Büttiker and Landauer [9, 10].

Deep impurity centres play a dominant role in the electronic properties of semiconductor materials and have therefore become a focus of extensive investigation [11–17]. Deep centres usually determine the nonequilibrium carrier lifetimes by acting as centres of nonradiative recombination and thermal ionization. The effect of an electric field on thermal ionization and carrier trapping has been used to probe deep impurities. In particular, investigation of the ionization or capture in a strong electric field is actually the only way to find the parameters of the phonon-assisted transitions determining the nonradiative recombination rate. DLTS is also among the most extensively employed tools. Most of the parameters of deep centres, such as ionization energy, and nonradiative and radiative trapping cross-sections, were obtained using various modifications of DLTS. Direct application of strong static electric fields is usually complicated by the onset of field nonuniformities in the sample and quite frequently initiates avalanche breakdown. Using an electric field of high intensity, short laser pulses at terahertz frequencies permit the contactless and uniform application of strong electric fields. Despite the high radiation intensities involved, there is no or only insignificant heating of the electron gas or of the crystal lattice under these conditions [5]. This is the result of the extremely weak absorption of the terahertz radiation due to the low concentration of free carriers (the carriers are frozen out on the centres), as well as to the use of short nanosecond-range pulses which do not substantially perturb the phonon system.

The observation of tunnelling ionization of deep impurities by contactless application of a strong uniform electric field using short FIR laser pulses revealed a new method for probing deep centres in semiconductors. Laser pulses shorter than the nonequilibrium carrier lifetimes allow one to measure the Huang–Rhys parameters of electron–phonon interaction in the adiabatic approximation, the structure of the adiabatic potentials, and the trapping kinetics of nonequilibrium carriers [18, 19].

This article is organized in the following way. After the introduction in section 1, in section 2 the thermal ionization of deep impurities in the adiabatic approximation is briefly reviewed, showing the importance of defect tunnelling in the ionization process. Section 3 deals with a rigorous theory of tunnelling ionization in high-frequency electric fields. In section 4 the experimental technique is described. Finally, in section 5 the experimental results are presented and discussed in relation to the theoretical background. It is shown that impurity ionization in a high-frequency electric field has been observed in a large variety of different defects and semiconductor hosts.

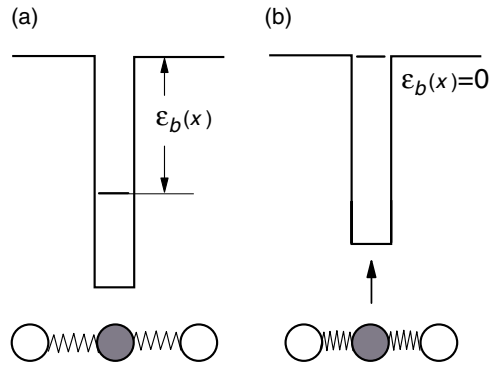

**Figure 1.** A schematic representation of the modulation of the impurity binding energy $\varepsilon_b$ by lattice vibrations; (a) the electronic ground state; (b) the electron merging into the continuum.

## 2. Thermal ionization of deep impurities

### 2.1. Adiabatic approximation

The binding energy of deep centres is much larger than the average phonon energy and therefore thermal emission and capture of carriers may only be achieved by involving many phonons. Since electronic transitions occur much faster than transitions in the phonon system, the adiabatic approximation can be used [20] and the electron–phonon interaction can be treated in the semiclassical model of adiabatic potentials. For the sake of simplicity and to be specific, we will discuss in the following electrons only, though the experimental investigation involves either electrons or holes.

We consider the simplest case of deep impurities having only one bound state. Obviously, this model applies directly to the emission and capture of carriers by neutral centres. However, as will be shown, the conclusions reached here remains valid also for deep impurities with attractive Coulomb potentials. The depth of the potential well depends sensitively on the separation of the impurity and the neighbouring atoms. Thus, vibrations of the impurity and lattice vibrations involving these atoms modulate the energy level of the impurity bound state [21], as sketched in figure 1. In the course of thermal vibrations the bound state level may eventually come up to the level of the continuous spectrum, enabling the electron to move from the localized state into the corresponding band, leaving the impurity in an ionized state or, more generally, in an electron-detached state. To describe this behaviour, a one-mode model with a single configuration coordinate $x$ is assumed. This approximation is justified because the breathing mode of local vibrations is most effective in phonon-assisted ionization and capture of deep impurities. In the adiabatic approximation, electronic transitions are assumed to occur at a constant configuration coordinate $x$. The vibrations of the impurity are determined by the potential due to the interaction with the surrounding atoms and due to the mean polarization field induced by the localized electron. Such a potential averaged over the electronic motion is called adiabatic. The magnitude of the potential includes the energy of the electron at a fixed coordinate $x$.

In figure 2 two basically different adiabatic potential diagrams are shown, representing an impurity with strong electron–phonon interaction and autolocalization, as used to describe the properties of DX and EL2 centres in III–VI compound semiconductors (figure 2(a)) and a substitutional on-site impurity of weak electron–phonon coupling (figure 2(b)). The potential curves $U_1(x)$ and $U_2(x)$ correspond to the electron bound to the impurity and to the ionized impurity with zero kinetic energy of the electron, respectively. The equilibrium position of the bound state is shifted with respect to the ionized state due to the electron–phonon interaction. The energy separation between the two potentials is determined by the electron binding energy $\varepsilon_b(x)$ as a function of the configuration coordinate $x$:

$$U_1(x) = U_2(x) - \varepsilon_b(x). \tag{1}$$

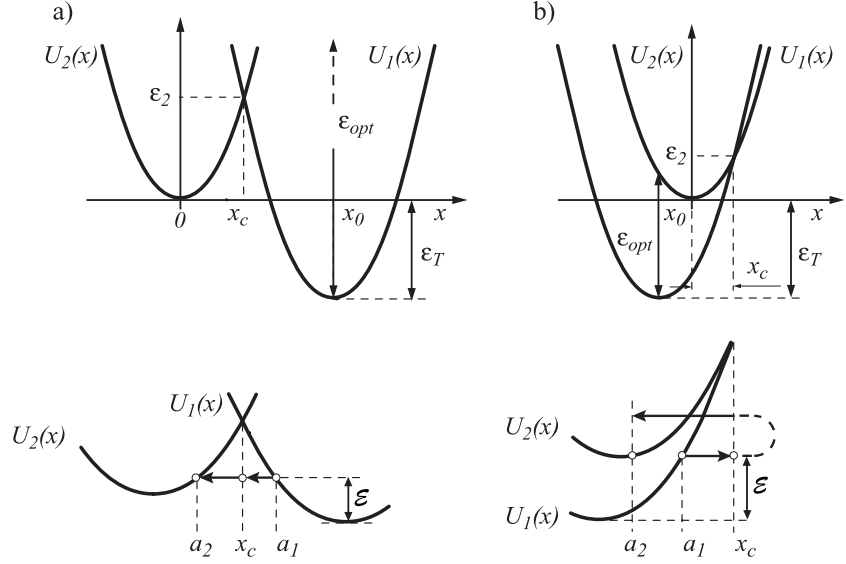

**Figure 2.** Upper diagrams: adiabatic potentials as a function of the configuration coordinate $x$ of the impurity motion for two possible schemes: (a) strong electron–phonon coupling with autolocalization; (b) weak electron–phonon coupling of substitutional impurities. $\varepsilon_T$ and $\varepsilon_{opt}$ are the thermal and optical activation energies, respectively. Solid curves $U_1(x)$ and $U_2(x)$ correspond to the carrier bound to the centre and detached from the impurity at the bottom of the band ($\varepsilon = 0$), respectively. Bottom diagrams: blown-up representations of the tunnelling trajectories.

Taking into account the Franck–Condon principle, the bound state equilibrium energy yields the value of the threshold of optical ionization: $\varepsilon_{opt} = \varepsilon_b (x = x_0)$, where $x_0$ is the displacement of the bound state due to electron–phonon interaction (see figure 2). Assuming the simple parabolic approach for $U_1(x)$, as shown in figure 2, $\varepsilon_{opt}$ is larger than the energy of thermal ionization $\varepsilon_T$, where $\varepsilon_T$ is the distance between the minima of the parabola. The relaxation energy $\Delta\varepsilon = \varepsilon_{opt} - \varepsilon_T$ characterizes the strength of the electron–phonon coupling. The larger the magnitude of the coupling, the larger $\Delta\varepsilon$. The electron–phonon coupling can be conveniently characterized by a dimensionless parameter

$$\beta = \frac{\Delta\varepsilon}{\varepsilon_T} = \frac{\varepsilon_{opt} - \varepsilon_T}{\varepsilon_T}. \tag{2}$$

The configuration of figure 2(a) illustrates the case of $\beta > 1$, where the optical and thermal ionization energies differ considerably. This diagram is used to describe, for instance, the DX and EL2 centres, where this difference was experimentally revealed [11–17]. Such autolocalized states have a large potential barrier suppressing the return of free carriers to the localized state, thus giving rise to the phenomenon of persistent photoconductivity. In these conditions, there is no radiative capture into the impurity state.

The configuration of figure 2(b) corresponds to weak electron–phonon coupling ($\beta < 1$). In this case the difference between $\varepsilon_{opt}$ and $\varepsilon_T$ is usually small, being several milli-electron volts. In fact for deep impurities in Ge and Si, a difference between $\varepsilon_{opt}$ and $\varepsilon_T$ has been observed only recently, by electric field-enhanced tunnelling ionization, as reviewed here [18]. There are, however, some cases where the relaxation energy $\Delta\varepsilon$ is large, as shown by Henry and Lang for 'state 2' oxygen in GaP where $\beta = 0.56$ [21]. The large value of $\Delta\varepsilon$ in this case has been attributed to the two different vibrational frequencies in the adiabatic potential of localized electrons and in the electron-detached state.

## 2.2. Thermal tunnelling emission and capture

The various features of the adiabatic potential configuration are of great importance for the thermal emission and the nonradiative capture of free carriers [14]. We shall restrict ourselves to the simple model of two identical displaced parabolic curves, which was first proposed by Huang and Rhys [20] and is currently widely employed in the theory of phonon-assisted transitions. By this model,

$$U_1(x) = \frac{M\omega_{vib}^2(x-x_0)^2}{2} - \varepsilon_T \tag{3}$$

$$U_2(x) = \frac{M\omega_{vib}^2 x^2}{2} \tag{4}$$

where $M$ and $\omega_{vib}$ are the mass of the impurity and the vibrational frequency, respectively.

Here we consider the zero-field electron emission from deep centres in equilibrium, where the emission rate is balanced by electron capture. The emission rate is identical to the capture rate. In a classical approach, the thermal emission probability is given by

$$e = \frac{2\pi}{\omega_{vib}} \exp\left(-\frac{\varepsilon_T + \varepsilon_2}{k_B T}\right) \tag{5}$$

where $\varepsilon_2 = U_1(x_c)$, and $x_c$ is the coordinate of the crossing of the potentials $U_1(x)$ and $U_2(x)$, at which the electron binding energy vanishes, $\varepsilon_b(x_c) = 0$ (see figure 2). Thus $\varepsilon_T + \varepsilon_2$ is the minimum excitation energy required to drive the electron into the continuum across the potential barrier separating $U_1(x)$ and $U_2(x)$. Adopting the Huang–Rhys model (equations (3), (4)) we get $\varepsilon_2 = (\varepsilon_T - \Delta\varepsilon)^2/4\Delta\varepsilon$. Usually the experimentally observed activation energy is much less than $\varepsilon_T + \varepsilon_2$. In fact the electron is emitted from a vibrational energy level $\mathcal{E}$ above the minimum of the potential $U_1(x)$ with $\varepsilon_T < \mathcal{E} \ll \varepsilon_T + \varepsilon_2$ (see figure 2). This is because the defect tunnels from the configuration corresponding to the electron bound state to that of the ionized impurity state or electron-detached state in the case of DX centres. As the vibrational energy $\mathcal{E}$ increases, the tunnelling barrier separating $U_1(x)$ and $U_2(x)$ becomes lower, and, hence, the tunnelling probability increases. On the other hand, the population of the energy level $\mathcal{E}$ decreases with increasing $\mathcal{E}$ proportionally to $\exp(-\mathcal{E}/k_B T)$. Thus for each temperature an optimum energy $\mathcal{E} = \mathcal{E}_m$ exists where the probability for tunnelling assumes a maximum [14, 15, 22, 23].

The defect tunnelling process will be treated in the semiclassical approximation. In this approach the particle has a well defined trajectory even under the potential barrier where the kinetic energy is negative. In this case the thermal emission of carriers can be described by a two-step process:

(i) Thermal excitation drives the vibrational system to an energy level $\mathcal{E} \geqslant \varepsilon_T$ in the bound state potential $U_1(x)$. The probability of this process is $P_T(\mathcal{E}) \propto \exp(-\mathcal{E}/kT)$.
(ii) A tunnelling reconstruction of the vibrational system occurs corresponding to a tunnelling at energy $\mathcal{E}$ from the bound potential $U_1(x)$ to the ionized potential $U_2(x)$ with the tunnelling probability $P_d(\mathcal{E})$ (see figure 2, lower diagrams). This process will be called defect tunnelling, in contrast to electron tunnelling which will be important in the electric field-enhanced tunnelling emission.

The ionization probability of the total process $P(\mathcal{E})$ is the given by

$$P(\mathcal{E}) = P_T(\mathcal{E}) P_d(\mathcal{E}). \tag{6}$$

The probability $P_d(\mathcal{E})$ of the tunnelling reconstruction of the vibrational system in the semiclassical approximation depends exponentially on the imaginary part of the action integral $S(\mathcal{E})$ multiplied by $1/\hbar$ and evaluated along the trajectory of tunnelling:

$$P_d(\mathcal{E}) \propto \exp(-2S(\mathcal{E})). \tag{7}$$

The total emission probability is then

$$P(\mathcal{E}) \propto \exp(-\psi) \tag{8}$$

with

$$\psi(\mathcal{E}) = \frac{\mathcal{E}}{k_B T} + 2S(\mathcal{E}). \tag{9}$$

The first term in equation (9) describes the population of the vibrational energy level $\mathcal{E}$, and the second, the defect tunnelling from the bound state to the electron-detached state. The optimum tunnelling energy $\mathcal{E}_m$ is determined by the vibrational energy at which $\Psi(\mathcal{E})$ has a minimum:

$$\left.\frac{d\psi(\mathcal{E})}{d\mathcal{E}}\right|_{\mathcal{E}=\mathcal{E}_m} = 2\left.\frac{dS(\mathcal{E})}{d\mathcal{E}}\right|_{\mathcal{E}=\mathcal{E}_m} + \frac{1}{k_B T} = 0. \tag{10}$$

The derivative $dS(\mathcal{E})/d\mathcal{E}$ in equation (10) multiplied by $-\hbar$ may be identified as the Büttiker–Landauer time $\tau$ for defect tunnelling [9, 10, 24] through the barrier at the optimum tunnelling energy $\mathcal{E}_m$:

$$\tau = -\hbar \left.\frac{dS_i(\mathcal{E})}{d\mathcal{E}}\right|_{\mathcal{E}=\mathcal{E}_m}. \tag{11}$$

Thus, in the case of phonon-assisted tunnelling ionization the time for tunnelling along the optimum trajectory is $\tau = \hbar/2k_B T$, determined only by the temperature.

The tunnelling trajectories for both adiabatic potential configurations are indicated in figure 2 (lower diagrams) by arrows. The trajectories start at the turning point $a_1$ and go under the potential $U_1(x)$ to the crossing point of the two adiabatic potentials $x_c$ and then to the turning point $a_2$ under potential $U_2$. Thus, after [14, 22, 23], $S(\mathcal{E})$ can be split into two parts in the form

$$S(\mathcal{E}) = -S_1(\mathcal{E}) + S_2(\mathcal{E}), \tag{12}$$

with

$$S_i(\mathcal{E}) = \frac{\sqrt{2M}}{\hbar} \int_{a_i}^{x_c} dx \sqrt{U_i(x) - (\mathcal{E} - \varepsilon_T)}, \qquad i = 1, 2. \tag{13}$$

The actual direction of tunnelling along the $x$-coordinate is specified by the sign of $S_i(\mathcal{E})$ in equation (13). Following from the orientation of the tunnelling trajectories, $S_1(\mathcal{E})$ is positive for the case of $\beta < 1$ (figure 2(b)) and negative for $\beta > 1$ (autolocalization; see figure 2(a)) because in this case $x_c < a_1$. The action integral $S_2(\mathcal{E})$ is positive for both adiabatic potential configurations.

The tunnelling times $\tau_1$ and $\tau_2$ under the corresponding adiabatic potentials are

$$\tau_i(\mathcal{E}_m) = -\hbar \left.\frac{dS_i(\mathcal{E})}{d\mathcal{E}}\right|_{\mathcal{E}=\mathcal{E}_m} = -\sqrt{\frac{M}{2}} \int_{a_i}^{x_c} \frac{dx}{\sqrt{U_i(x) - (\mathcal{E}_m - \varepsilon_T)}} \qquad i = 1, 2. \tag{14}$$

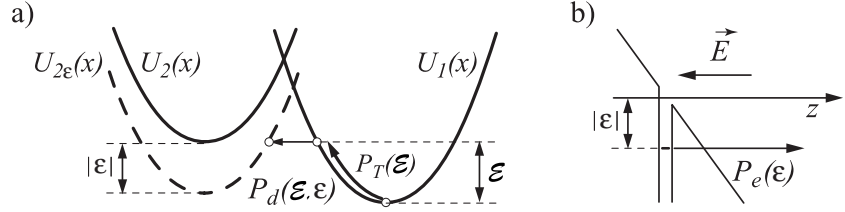

**Figure 3.** Illustrations of the tunnelling ionization of deep centres in a static electric field: (a) thermal excitation and defect tunnelling; (b) electron tunnelling with probabilities $P_T(\mathcal{E})$, $P_d(\mathcal{E}, \varepsilon)$, and $P_e(\varepsilon)$.

One can see that they are given by integration over the distance of tunnelling divided by the magnitude of the velocity under the barrier.

Equations (10)–(14) yield

$$\tau = \tau_2 \pm |\tau_1| = \frac{\hbar}{2k_B T} \quad (15)$$

where the minus and plus signs correspond to the configurations of figures 2(a) and (b), respectively. Since $(\mathcal{E}_m - \varepsilon_T)$ is usually much smaller than $\varepsilon_T$, the time $\tau_1$ is practically temperature independent and can be calculated for $\mathcal{E}_m = \varepsilon_T$.

## 3. Theory of electric field-enhanced tunnelling ionization

### 3.1. Ionization probability

Carrier emission in static electric fields was first considered in [25] and calculated numerically in [26]; analytical expressions for the probability of deep-impurity-centre ionization were obtained in [27, 28].

In a homogeneous electric field a potential of constant slope along the direction of the field vector is superimposed on the potential well binding the electron to the impurity. A triangular potential barrier is formed, that the electron may cross by tunnelling on a level of negative kinetic energy $\varepsilon$. The adiabatic potential of the unbound state is thus shifted down in energy to

$$U_{2\varepsilon}(x) = U_2(x) + \varepsilon \quad (\varepsilon < 0) \quad (16)$$

(dashed lines in figures 3(a) and 4) shortening the defect tunnelling trajectory in configuration space and lowering the barrier height. Thus the electron emission is enhanced above the level of thermal ionization in equilibrium. In alternating electric fields of high-frequency radiation, the electrons predominantly leave the impurities if the electric field assumes its peak amplitude. One of the main issues of the following discussions will be the question of how far this model applies with increasing frequency.

The electric field-stimulated emission of carriers consists of three simultaneously proceeding processes: the two processes like in the case of thermal equilibrium emission of carriers, namely:

(i) thermal excitation of the vibrational system and
(ii) the tunnelling reconstruction of the vibrational system (illustrated in figure 3(a) for the case of strong electron–phonon interaction); and in addition a new one:
(iii) tunnelling of the electron through the triangular potential formed by the attractive force of the impurity and the electric field (figure 3(b)).

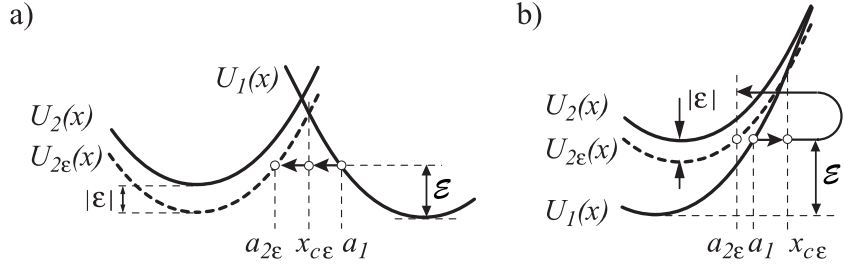

**Figure 4.** Blown-up representations of the tunnelling trajectories for (a) strong electron–phonon coupling (autolocalization) and (b) weak electron–phonon coupling. The dashed curves show the potential $U_{2\varepsilon}(x)$ of the system: an ionized impurity and an electron with negative kinetic energy $\varepsilon$ obtained by electron tunnelling in an electric field.

Electron tunnelling occurs at an energy $\varepsilon < 0$ with the probability $P_e(\varepsilon)$. The electric field acts on electron tunnelling only, and the stimulation of thermal tunnelling ionization of impurities is caused by the lowering of the ionized adiabatic potential from $U_2(x)$ to $U_{2\varepsilon}(x)$ with $\varepsilon < 0$ (figures 3(a) and 4). The ionization probability of the total process depends now also on the electron energy $\varepsilon$ and is given by

$$P(\mathcal{E}, \varepsilon) = P_T(\mathcal{E}) P_d(\mathcal{E}, \varepsilon) P_e(\varepsilon). \tag{17}$$

The probability $P_d(\mathcal{E}, \varepsilon)$ of the tunnelling reconstruction of the vibrational system in the semiclassical approximation is calculated similarly to in the case of thermal emission. The tunnelling trajectories and potential barriers for weak and strong electron–phonon interaction are shown in figure 4.

The trajectories, as in the case of thermal tunnelling ionization, are split into two parts, both under barriers formed by the potentials $U_1(x)$ and here $U_{2\varepsilon}(x)$. The system tunnels at energy level $\mathcal{E} - \varepsilon_T$ from the turning point $a_{1\varepsilon}$ to $x_{c\varepsilon}$ under the potential $U_1(x)$ and from $x_{c\varepsilon}$, where $x = x_{c\varepsilon}$ is the intersection of the potentials, to the turning point $a_{2\varepsilon}$ under $U_{2\varepsilon}(x)$ (equation (16)). We note that all energies are measured from the bottom of the potential $U_2(x)$. The tunnelling probability depends exponentially on the imaginary part of the principal function (action) evaluated along the trajectory linking the turning points $a_1$ and $a_{2\varepsilon}$. Then the probability of defect tunnelling is given by

$$P_d(\mathcal{E}, \varepsilon) \propto \exp(-2(S_{2\varepsilon}(\mathcal{E}, \varepsilon) - S_{1\varepsilon}(\mathcal{E}, \varepsilon))) \tag{18}$$

where

$$S_{1\varepsilon}(\mathcal{E}, \varepsilon) = \frac{\sqrt{2M}}{\hbar} \int_{a_1}^{x_{c\varepsilon}} \sqrt{U_1(x) - (\mathcal{E} - \varepsilon_T)} \, dx \tag{19}$$

and

$$S_{2\varepsilon}(\mathcal{E}, \varepsilon) = \frac{\sqrt{2M}}{\hbar} \int_{a_{2\varepsilon}}^{x_{c\varepsilon}} \sqrt{U_{2\varepsilon}(x) - (\mathcal{E} - \varepsilon_T)} \, dx. \tag{20}$$

The probability of electron tunnelling in the alternating electric field $P_e(\varepsilon)$ is calculated semiclassically using the short-radius potential model for impurities after [24, 29]. The unperturbed wavefunction $\Psi_0(r, t)$ of an electron in a short-range potential at $r = 0$ on the energy level $\varepsilon$ ($\varepsilon < 0$) is given by

$$\Psi_0(r = 0, t) \propto \exp(-i\varepsilon t/\hbar). \tag{21}$$

The electron wavefunction in an arbitrary point $r$ and for the time $t$ is

$$\Psi(r, t) \propto e^{(i/\hbar)\tilde{S}(r,t)}, \tag{22}$$

where $\tilde{S}(r, t)$ is the electron action. In order to determine $\tilde{S}(r, t)$ as a function of the electron coordinates $r$ and the time $t$, one should find the general integral [30] of the Hamilton–Jacobi equations

$$\frac{\partial \tilde{S}}{\partial t} = -\mathcal{H}(p, r, t); \qquad \nabla \tilde{S} = p \qquad (23)$$

which depends on an arbitrary function. Here $\mathcal{H}$ and $p$ are the Hamiltonian and the electron momentum, respectively. The resulting principal function $\tilde{S}(r, t)$ can be written in the form

$$\tilde{S} = \tilde{S}_0 - \varepsilon t_0; \qquad \tilde{S}_0 = \int_{t_0}^{t} \mathcal{L}(r', \dot{r}', t') \, dt', \qquad (24)$$

where $\mathcal{L}(r', \dot{r}', t')$ is the Lagrange function. The position vector $r'(t')$ as a function of $t'$ can be found by solving the classical equation of motion with boundary conditions

$$r'(t')|_{t'=t_0} = 0; \qquad r'(t')|_{t'=t} = r. \qquad (25)$$

The principal function is obtained in the form of equation (24) by taking it into account that the wavefunction $\Psi(r, t)$ is equal to the unperturbed wavefunction $\Psi_0(r, t)$ at $r = 0$ (at the defect) and that $r = 0$ at $t = t_0$ (equation (25)). In equations (24) and (25), $t_0$ is a function of $r$ and $t$ which should be found from the equation

$$\left(\frac{\partial \tilde{S}}{\partial t_0}\right)_{r,t} = 0. \qquad (26)$$

We want to emphasize that $r$ and $t$ are real while $r'$, $t'$, and $t_0$ can be complex. As will be shown below, the imaginary part of $t_0$ determines the Büttiker–Landauer electron tunnelling time and is a function of the electron energy $\varepsilon$.

The electron tunnelling probability $P_e(\varepsilon)$, being determined by the current density flowing from the centre, is proportional to $|\Psi|^2$ in the region of $r$ outside the potential well where the electron is free. To find $P_e(\varepsilon)$, it is sufficient to calculate Im $\tilde{S}$ in the vicinity of its maximum, i.e. at values of $r$ where

$$\text{Im } \nabla \tilde{S} = \text{Im}(p|_{t'=t}) = 0. \qquad (27)$$

For this region of space, it follows that, after the left side of equations (23),

$$\frac{\partial (\text{Im } \tilde{S})}{\partial t} = 0. \qquad (28)$$

Thus, the probability of electron tunnelling $P_e(\varepsilon)$ can be written as

$$P_e(\varepsilon) = \exp(-2S_e(\varepsilon)) \qquad (29)$$

where

$$S_e(\varepsilon) = \frac{\text{Im } \tilde{S}}{\hbar}. \qquad (30)$$

Here $\tilde{S}$ is determined by equations (24), (26), and (27). Note that in calculating $S_e(\varepsilon)$, one can arbitrarily take a value of the time $t$ according to equation (28); we will assume $t = 0$.

The electron tunnelling time $\tau_e(\varepsilon)$ is determined as

$$\tau_e(\varepsilon) = -\hbar \frac{\partial S_e(\varepsilon)}{\partial \varepsilon}. \qquad (31)$$

Using equations (24) and (30) we obtain

$$\tau_e(\varepsilon) = -\text{Im} \frac{\partial \tilde{S}}{\partial \varepsilon} = -\text{Im}\left(-t_0 + \frac{\partial \tilde{S}}{\partial t_0} \frac{\partial t_0}{\partial \varepsilon}\right). \qquad (32)$$

According to equation (26) we finally get

$$\tau_e(\varepsilon) = \operatorname{Im} t_0. \tag{33}$$

As a result, the probability of ionization $e(E)$ as a function of the electric field $E$ is obtained by integrating equation (17) over $\mathcal{E}$ and $\varepsilon$:

$$e(E) = \int\int P_e(\varepsilon) P_d(\mathcal{E},\varepsilon) \exp(-\mathcal{E}/k_B T) \, d\varepsilon \, d\mathcal{E}. \tag{34}$$

Calculating this integral by the saddle point method, we see that there is a vibrational energy $\mathcal{E} = \mathcal{E}_m$ and an electron energy $\varepsilon = \varepsilon_m$ where the ionization probability has a sharp maximum. Thus, defect and electron tunnelling take place mostly at these energy levels and the ionization probability can be written in the following approximate form:

$$e(E) \propto P_e(\varepsilon_m) P_d(\mathcal{E}_m, \varepsilon_m) \exp(-\mathcal{E}_m/k_B T). \tag{35}$$

The defect and the electron tunnelling at the energy levels $\mathcal{E}_m$ and $\varepsilon_m$ can be characterized by a defect tunnelling time $\tau$ and an electron tunnelling time $\tau_e \equiv \tau_e(\varepsilon_m)$, respectively. The saddle point method applied to equation (34) yields that the defect tunnelling time is determined by the temperature [6]:

$$\tau = \tau_{2\varepsilon}(\mathcal{E}_m, \varepsilon_m) - \tau_{1\varepsilon}(\mathcal{E}_m, \varepsilon_m) = \frac{\hbar}{2k_B T} \tag{36}$$

where $\tau_{n\varepsilon}(\mathcal{E}_m, \varepsilon_m)$ are times for tunnelling under the barriers $U_{n\varepsilon}(x)$ of the vibrational system with

$$\tau_{n\varepsilon}(\mathcal{E},\varepsilon) = -\hbar\frac{\partial S_{n\varepsilon}(\mathcal{E},\varepsilon)}{\partial \mathcal{E}} \qquad (n=1,2). \tag{37}$$

As an important result obtained by the saddle point method for solving the integral equation (34) is that the electron tunnelling time $\tau_e(\varepsilon_m)$ is equal to the defect tunnelling time $\tau_{2\varepsilon}(\mathcal{E}_m, \varepsilon_m)$ under the potential $U_{2\varepsilon}(x)$ of the ionized configuration:

$$\tau_e(\varepsilon_m) = \tau_{2\varepsilon}(\mathcal{E}_m, \varepsilon_m). \tag{38}$$

The solution of equations (36) and (38) allows one to find $\mathcal{E}_m$ and $\varepsilon_m$.

### 3.2. Phonon-assisted tunnelling

The tunnelling ionization probability in the limit of not-too-high electric fields and not-too-low temperatures is dominated by phonon-assisted tunnelling. The electric field and temperature limits will be defined more precisely below. The theory is developed for neutral impurities, which means that the Coulomb force between the carrier and the centre is ignored when the carrier is detached from the impurity centre. The tunnelling ionization of charged impurities in static and alternating electric fields [19] will be discussed below, where we show that at low electric field strengths, ionization is caused by the Poole–Frenkel effect, whereas at high fields, tunnelling ionization enhanced by the Coulomb force dominates the emission process. In the case of the phonon-assisted tunnelling the optimum electron tunnelling energy $\varepsilon_m$ is small in comparison to the optimum defect tunnelling energy $\mathcal{E}_m$. In this limit the tunnelling emission probability $e(E)$ of carriers can be calculated analytically. The effect of the electric field is a small shift of the ionized potential $U_2(x)$ to a lower level $U_{2\varepsilon}(x)$. The potential $U_1(x)$ is not affected by the electric field. For small $\varepsilon$ the quantities $S_{1\varepsilon}(\mathcal{E},\varepsilon)$ and $S_{2\varepsilon}(\mathcal{E},\varepsilon)$ can be taken into account in the linear approximation as a function of $\varepsilon$. Then we obtain $S_{2\varepsilon}(\mathcal{E},\varepsilon) - S_{1\varepsilon}(\mathcal{E},\varepsilon) = S_2(\mathcal{E},\varepsilon) - S_1(\mathcal{E},\varepsilon) + \tau_2\varepsilon/\hbar$, where $S_1(\mathcal{E},\varepsilon)$, $S_2(\mathcal{E},\varepsilon)$, $\tau_2$ are calculated after equations (19), (20), and (37) for $\varepsilon = 0$ and are independent of the electric field. Taking into account equations (31)–(36) and (33), we find the dependence of the ionization probability on the electric field in the form

$$e(E) = e(0) \exp\left(-\frac{2}{\hbar} \operatorname{Im} \tilde{S}_0(\varepsilon_m)\right) \tag{39}$$

where $e(0)$ is the thermal ionization probability at zero field; $\tilde{S}_0$ follows from equation (24) and should be calculated in the range where equation (27) is satisfied at arbitrary $t$.

*3.2.1. Frequency dependence.* Now we consider the frequency dependence of phonon-assisted tunnelling and determine $\operatorname{Im} \tilde{S}_0(\varepsilon_m)$. If an alternating electric field $\boldsymbol{E}(t)$ is applied to an electron, the Lagrange function has the form

$$\mathcal{L}(\boldsymbol{r}', \dot{\boldsymbol{r}}', t') = \frac{m\dot{r}'^2}{2} + e\,(\boldsymbol{r}' \cdot \boldsymbol{E}(t')) \tag{40}$$

where $m$ and $e$ are the mass and the charge of the electron, respectively. The equation of motion is

$$m\ddot{\boldsymbol{r}}' = e\boldsymbol{E}(t'). \tag{41}$$

Integrating equation (24) by parts and taking into account equations (40) and (41) with the boundary conditions given by the equations (25) and (27), we get

$$\operatorname{Im} \tilde{S}_0 = -\operatorname{Im} \int_{t_0}^{t} \frac{m\dot{r}'^2}{2} \, dt'. \tag{42}$$

In the following we consider the general case of elliptically polarized radiation of frequency $\omega$:

$$E_x = E_1 \cos \omega t, \qquad E_y = E_2 \sin \omega t \tag{43}$$

propagating in the $z$-direction. Then we have from equation (41)

$$\dot{x}' = \frac{eE_1}{m\omega} \sin \omega t' + u_x \tag{44}$$

$$\dot{y}' = -\frac{eE_2}{m\omega} \cos \omega t' + u_y \tag{45}$$

and hence

$$x' = -\frac{eE_1}{m\omega^2}(\cos \omega t' - \cos \omega t_0) + u_x(t' - t_0) \tag{46}$$

$$y' = -\frac{eE_2}{m\omega^2}(\sin \omega t' - \sin \omega t_0) + u_y(t' - t_0). \tag{47}$$

Here the velocities $u_x$ and $u_y$ are real constants (see equation (27)) which are related to the velocity $\boldsymbol{v_0} = \dot{\boldsymbol{r}}(t' = t_0)$ by the equations

$$u_x = -\frac{eE_1}{m\omega} \sin \omega t_0 + v_{0x} \tag{48}$$

$$u_y = \frac{eE_2}{m\omega} \cos \omega t_0 + v_{0y}. \tag{49}$$

According to equation (33), the complex time $t_0$ can be written in the form

$$t_0 = t'_0 + i\tau_e. \tag{50}$$

Taking into account that

$$\left.\frac{\partial \tilde{S}_0}{\partial t_0}\right|_{r,t} = H(\boldsymbol{r}', \boldsymbol{t}')|_{t'=t_0, r'=0} \tag{51}$$

we have from equation (26)

$$\varepsilon = \frac{mv_0^2}{2} \tag{52}$$

where $\varepsilon < 0$ and hence $v_0^2$ is negative (see [30]).

According to equations (48), (50), and (52), we get

$$\frac{2\varepsilon}{m} = \left(u_x + \frac{eE_1}{m\omega}(\cosh \omega\tau_e(\varepsilon) \sin \omega t'_0 + i \sinh \omega\tau_e(\varepsilon) \cos \omega t'_0)\right)^2$$
$$+ \left(u_y - \frac{eE_2}{m\omega}(\cosh \omega\tau_e(\varepsilon) \cos \omega t'_0 + i \sinh \omega\tau_e(\varepsilon) \sin \omega t'_0)\right)^2. \tag{53}$$

The condition that the coordinates $x'$ and $y'$ should be real at $t' = t$, together with equation (50), gives

$$u_x = -\frac{eE_1}{m\omega^2 \tau_e(\varepsilon)} \sinh \omega\tau_e(\varepsilon) \sin \omega t'_0 \tag{54}$$

$$u_y = \frac{eE_2}{m\omega^2 \tau_e(\varepsilon)} \sinh \omega\tau_e(\varepsilon) \cos \omega t'_0. \tag{55}$$

Now we can see that the left part of equation (53) can be real only if $\sin \omega t'_0 = 0$ or $\cos \omega t'_0 = 0$. For $E_2 = 0$ (linear polarization), $\cos \omega t'_0$ cannot be equal to zero, as it leads to positive values of $\varepsilon$, while they must be negative due to electron tunnelling. Thus, only the first situation is practically realized. In this case we have from equation (53) a relation between the tunnelling time $\tau_e$ and the electron energy $\varepsilon$:

$$\frac{2\varepsilon}{m} = -\left(\frac{eE_1}{m\omega}\right)^2 \sinh^2(\omega\tau_e(\varepsilon)) + \left(\frac{eE_2}{m\omega}\right)^2 \left(\frac{\sinh \omega\tau_e(\varepsilon)}{\omega\tau_e(\varepsilon)} - \cosh \omega\tau_e(\varepsilon)\right)^2. \tag{56}$$

Then, after integration of equation (42), an expression for $\operatorname{Im} \tilde{S}_0$ is obtained:

$$\operatorname{Im} \tilde{S}_0 = \frac{e^2 \tau_e(\varepsilon)}{4m\omega^2}\left[E_1^2\left(1 - \frac{\sinh 2\omega\tau_e(\varepsilon)}{2\omega\tau_e(\varepsilon)}\right) + E_2^2\left(1 + \frac{\sinh 2\omega\tau_e(\varepsilon)}{2\omega\tau_e(\varepsilon)} - 2\frac{\sinh^2 \omega\tau_e(\varepsilon)}{(\omega\tau_e(\varepsilon))^2}\right)\right]. \tag{57}$$

In the case of phonon-assisted tunnelling considered here, the electron tunnelling time $\tau_e(\varepsilon_m)$ is equal to the defect tunnelling time $\tau_2$, because $\tau_e(\varepsilon_m) = \tau_{2\varepsilon}(\mathcal{E}_m, \varepsilon_m)$ (equation (38)) and $\tau_{2\varepsilon}(\mathcal{E}_m, \varepsilon_m)$ is equal to the tunnelling time $\tau_2$ (equation (15)) for $|\varepsilon_m| \ll \mathcal{E}_m$.

Therefore, the probability of the phonon-assisted tunnelling can be obtained by using equations (39) and (57), replacing $\tau_e(\varepsilon_m)$ by $\tau_2$.

From equation (39) we get the ionization probability as a function of the electric field $E$:

$$e(E) = e(0) \exp\left(\frac{E^2}{E_c^2}\right). \tag{58}$$

It is convenient to write $E_c$ in the form

$$E_c^2 = \frac{3m\hbar}{e^2(\tau_2^*)^3} \tag{59}$$

introducing an effective time $\tau_2^*$. This time may be obtained from equation (57). In the case of linear polarization ($E_1 = E$, $E_2 = 0$), we find

$$(\tau_2^*)^3 = \frac{3\tau_2}{2\omega^2}\left(\frac{\sinh 2\omega\tau_2}{2\omega\tau_2} - 1\right), \tag{60}$$

and for circular polarization ($E_1 = E_2 = E$), we have

$$(\tau_2^*)^3 = \frac{3\tau_2}{\omega^2}\left(\frac{\sinh^2 \omega\tau_2}{(\omega\tau_2)^2} - 1\right). \tag{61}$$

These relations show that for a static electric field ($\omega = 0$) the effective time $\tau_2^*$ is equal to the defect tunnelling time $\tau_2$. Therefore, equations (58) and (59) are in agreement with previous derivations of the tunnelling emission probability in static fields [28].

The equations (58)–(61) have been obtained without any assumption about the shape of the adiabatic potentials $U_1(x)$ and $U_2(x)$. In fact, the defect tunnelling time $\tau = \hbar/2k_BT$ is independent of the form of the potentials. However, the parameters $\tau_1$ and $\tau_2 = \tau + \tau_1$, which are crucial for electric field-stimulated tunnelling, substantially depend on the configuration of the potentials. In the model of parabolic potentials (the Huang–Rhys model; see equations (3) and (4)), $\tau_1$ is given by

$$\tau_1 = \frac{1}{2\omega_{vib}} \ln \frac{\varepsilon_T}{\Delta \varepsilon}. \tag{62}$$

From equation (62) it follows that the tunnelling time $\tau_1$ is negative for autolocalized impurities with $\Delta \varepsilon > \varepsilon_T$ and positive for substitutional impurities with weak electron–phonon interaction where $\Delta \varepsilon < \varepsilon_T$. Thus we get

$$\tau_2 = \frac{\hbar}{2kT} \pm |\tau_1| \tag{63}$$

with the plus and minus signs for substitutional and autolocalized impurities, respectively. The tunnelling time $\tau_2$ controls defect tunnelling in static fields; it is a function of the temperature and the shape of the adiabatic potentials, and is independent of the frequency $\omega$. The effective time $\tau_2^*$ which controls tunnelling for all frequencies additionally depends on $\omega$. The frequency dependence of $\tau_2^*/\tau_2$ as function of $\omega \tau_2$ is displayed in figure 5 for linearly and circularly polarized radiation. As long as $\omega \tau_2 \leqslant 1$, the ratio $\tau_2^*/\tau_2$ is equal to one, and in this quasi-static regime the ionization probability is independent of the electric field frequency and the state of polarization. For $\omega \tau_2 > 1$ the ratio increases, drastically enhancing the ionization probability. In this high-frequency regime, the ionization probability is polarization dependent, being higher for linearly than circularly polarized radiation at the same amplitude $E$ of the electric field.

*3.2.2. Magnetic field dependence.* An external magnetic field $B$ applied perpendicularly to the electric field which generates the tunnelling barrier decreases the probability of electron tunnelling. This effect has been theoretically investigated for tunnelling of electrons through static potential barriers [29, 31]. The theory has been extended to phonon-assisted tunnelling ionization of deep impurities in dc electric fields [32] and in high-frequency alternating fields [33] showing that in the case of phonon-assisted tunnelling also, the carrier emission is suppressed by an external magnetic field ($\boldsymbol{B} \perp \boldsymbol{E}$). For the ionization probability we get again an exponential dependence on the square of the electric field strength $\propto \exp(E^2/E_c^2)$ where we write $E_c^2$ in the form of equation (59), defining by this an effective time $\tau_2^*$ which depends now on the magnetic field strength:

$$(\tau_2^*)^3 = \frac{3\omega_c^2}{(\omega^2 - \omega_c^2)^2} \left\{ \int_0^{\tau_2} \left[ \left( -\cosh \omega \tau + \frac{\omega_c}{\omega} \frac{\sinh \omega \tau_2}{\sinh \omega_c \tau_2} \cosh \omega_c \tau \right)^2 d\tau \right. \right.$$
$$\left. \left. + \int_0^{\tau_2} \left( \frac{\omega}{\omega_c} \sinh \omega \tau - \frac{\omega_c}{\omega} \frac{\sinh \omega \tau_2}{\sinh \omega_c \tau_2} \sinh \omega_c \tau \right)^2 \right] d\tau \right\}. \tag{64}$$

Here $\omega_c = eB/mc$ is the cyclotron frequency. In figure 6 the calculated $\tau_2^*$ normalized by the frequency-independent tunnelling time $\tau_2$ is shown as a function of $\omega \tau_2$ for different parameters $\omega_c \tau_2 \propto B$. The suppression of the tunnelling probability occurs in both frequency ranges: at low frequencies when tunnelling is independent of frequency as well as at high frequencies when the tunnelling probability increases drastically with rising frequency. The effect of a magnetic field on tunnelling is strongest if $\bm{B}$ is oriented normal to the tunnelling trajectory and vanishes if $\bm{B}$ is parallel to the electric field.

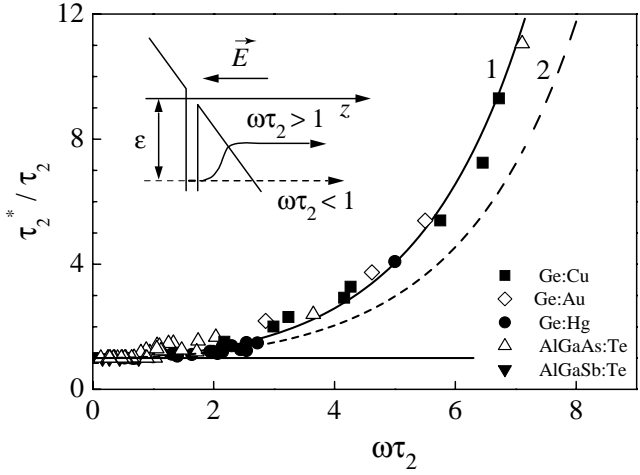

**Figure 5.** The ratio $\tau_2^*/\tau_2$ as a function of $\omega\tau_2$. The curves show the dependences calculated according to equation (60) for linear polarization (curve 1) and equation (61) for circular polarization (curve 2). Experimental results obtained with linearly polarized radiation are plotted for all materials, all temperatures, and all radiation frequencies of the present investigation. The inset shows the electron tunnelling trajectory in the quasi-static limit (broken curve) and the high-frequency regime (full curve).

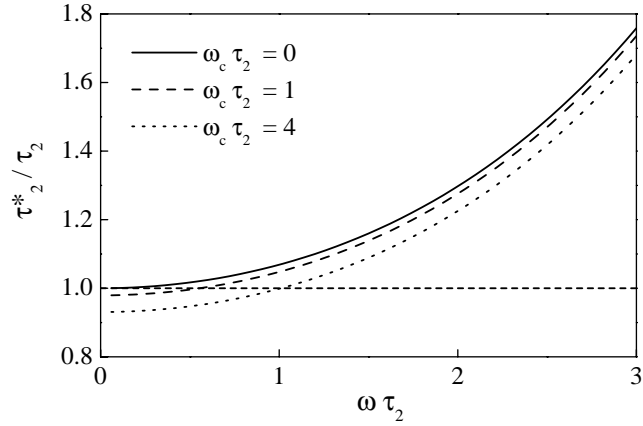

**Figure 6.** The ratio $\tau_2^*/\tau_2$ versus $\omega\tau_2$ calculated after equation (64) for various values of $\omega_c\tau_2$ and $\bm{B} \perp \bm{E}$.

### 3.3. Direct tunnelling ionization

The phonon-assisted tunnelling regime with the characteristic electric field dependence of the ionization probability $e(E) \propto \exp(E^2/E_c^2)$ is limited by the condition that the optimal electron tunnelling energy is smaller than the optimal defect tunnelling energy, $|\varepsilon_m| < \mathcal{E}_m$. For linear

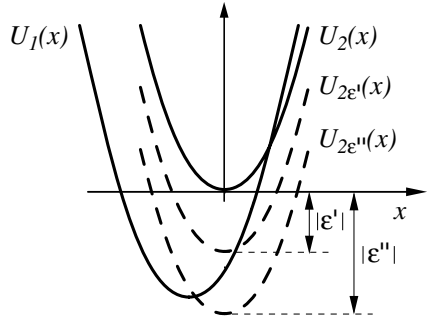

**Figure 7.** A schematic plot of the adiabatic potentials in high electric fields. With rising electric field strength, $|\varepsilon|$ increases; thus, $U_{2\varepsilon}(x)$ shifts to lower energy. The crossing point of $U_{2\varepsilon}(x)$ approaches the minimum of the bound state $U_1(x)$. The tunnelling process proceeds from phonon-assisted tunnelling to direct tunnelling at the crossing point without involving phonons.

polarization ($E_1 = E$ and $E_2 = 0$), this inequality can be written in the form

$$\frac{(eE\tau_e)^2}{2m} < \varepsilon_T \frac{(\omega\tau_e)^2}{\sinh^2(\omega\tau_e)}. \tag{65}$$

In this limit electron tunnelling yields only a small correction to the defect tunnelling, and the electron tunnelling time $\tau_e$ is equal to the defect tunnelling time $\tau_2$ and is independent on the electron energy $\varepsilon$. If the inequality is violated ($|\varepsilon_m| > \mathcal{E}_m$), direct tunnelling dominates the ionization process. Now the tunnelling times become dependent on the electron energy $\varepsilon$. This occurs at high electric fields, which shifts $|\varepsilon_m|$ to higher values. The adiabatic potential of the ionized defect for various field strengths is shown in figure 7. With rising electric field strength, the magnitude of the electron energy, $|\varepsilon|$, increases and the potential curve of the electron-detached state $U_{2\varepsilon}(x)$ is shifted to lower energy. The crossing point of the $U_{2\varepsilon}(x)$ and $U_1(x)$ decreases on the energy scale approaching the minimum of $U_1(x)$. Now direct electron tunnelling takes place at the crossing point of these potential curves without the assistance of phonons.

As equation (65) shows, the electric field strength where the transition from phonon-assisted tunnelling to direct tunnelling occurs decreases with increasing frequency and/or decreasing temperature because of the temperature dependence of $\tau_e$ (equations (38), (63)). In the regime of direct tunnelling, the ionization probability approaches the well known relation for electron tunnelling through a triangular barrier [24]. The effect of thermal phonons can be considered as a small perturbation which decays with rising electric field strength. The emission probability is found to be independent of frequency and can be written as

$$e(E) = \frac{eE}{2\sqrt{2m\varepsilon_{opt}}} \exp(-\phi) \tag{66}$$

with

$$\phi = \frac{4\sqrt{2m}}{3\hbar eE}\varepsilon_{opt}^{3/2} - b\frac{m\omega_{vib}\varepsilon_{opt}^2}{\hbar e^2 E^2}\coth\frac{\hbar\omega_{vib}}{2kT}. \tag{67}$$

Here $b$ is a constant. In the Huang–Rhys model, $b = 4\,\Delta\varepsilon/\varepsilon_{opt}$ where $\Delta\varepsilon = \varepsilon_{opt} - \varepsilon_T$. The first term in equation (67) is the exponent for electron tunnelling through a triangular barrier (see figure 3(b)) while the second term is a correction due to thermal phonons.

The deviation from phonon-assisted tunnelling at increasing electric field strength can be utilized to characterize deep impurities. The transition from phonon-assisted tunnelling to direct tunnelling strongly depends on $\Delta\varepsilon = \varepsilon_{opt} - \varepsilon_T$ and $\omega_{vib}$ and allows one to determine these parameters. To demonstrate this dependence, the emission probability of neutral impurities has been calculated for different combinations of these parameters at constant tunnelling time $\tau_1$, taking into account both phonon-assisted and direct tunnelling. The results are plotted in figure 8, showing that the electric field strength of the transition shifts to higher values with rising local vibration frequency.

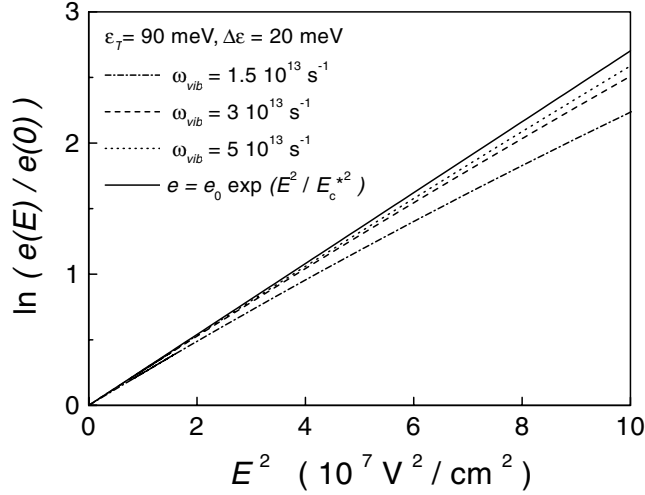

**Figure 8.** Calculations for ionization of the neutral impurity with $\varepsilon_T = 90$ meV and $m^* = 0.044\ m_e$. The ratio of the emission probability in the electric field $e(E)$ and the thermal emission probability $e(0)$ is plotted as a function of $E^2$ for different values of $\omega_{vib}$ and $\Delta\varepsilon$ at $\tau_1 = 2.9 \times 10^{-14}$ s.

*3.4. The effect of impurity charge*

Most deep centres bear a charge, which means that a Coulomb force is acting between the detached carrier and the impurity centre. Thus the long-range Coulomb potential must be taken into account in addition to the deep well causing the large binding energy of the carrier. For a Coulomb potential, in contrast to the narrow potential well, the height of an energy barrier formed by an external electric field is lowered along the direction of the electric field vector, as sketched in figure 9. Therefore an electric field yields an increase of the thermal emission probability by excitation of the carrier across the barrier, without tunnelling. This thermal ionization process is called the Poole–Frenkel effect [34, 35]. It has been observed in the current–voltage characteristics under dc conditions in many insulators and semiconductors. The Poole–Frenkel effect is the dominant mechanism of electric field-assisted thermal ionization at not-too-high field strengths before tunnelling of carriers becomes important [19].

A simple calculation shows that in an electric field $E$, the ionization barrier is diminished by an amount $\varepsilon_{PF}$: given by

$$\varepsilon_{PF} = 2\sqrt{\frac{Ze^3 E}{\kappa}} \tag{68}$$

where $Z$ is the charge of the centre and $\kappa$ is the dielectric constant.

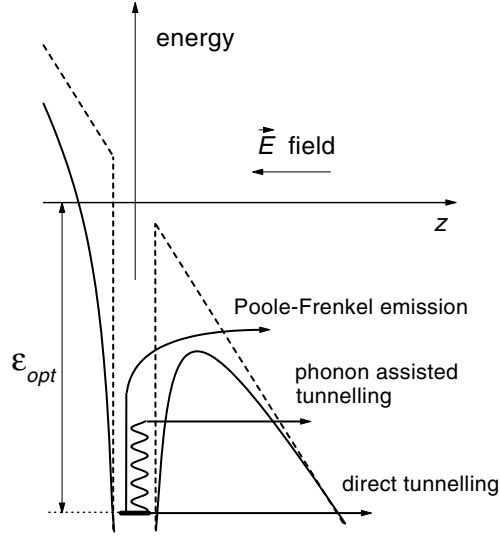

**Figure 9.** The potential of a charged deep impurity in the absence (a) and presence (b) of an electric field $E$ applied along axis $z$. $\varepsilon_b$ is the ground state binding energy and $\varepsilon_{PF}$ is the Poole–Frenkel reduction of the potential barrier height.

Owing to this fact, the probability of thermal emission due to an electric field increases as

$$e(E)/e(0) = \exp(\varepsilon_{PF}/kT). \tag{69}$$

In semiconductors this effect is observed for attractive Coulomb impurity centres at high temperatures and electric field strengths $E$ less than the field which yields $\varepsilon_{PF}(E) = Z^2\,\mathrm{Ryd}^\star$, where $\mathrm{Ryd}^\star$ is the effective Rydberg energy of the electron in the Coulomb potential of the charged impurity. The current flow in the sample in this case increases exponentially with the square root of the applied electric field.

There are, however, several disagreements between experiment and the Frenkel theory. In particular, experimental studies showed that the slope of $\ln e(E)/e(0)$ versus $E$ is only about one half of that derived from equations (68) and (69) and that at very low electric field strengths the emission rate becomes practically constant. These discrepancies are resolved by more realistic theoretical approaches which consider the emission of carriers in three dimensions [36, 37], take into account carrier distribution statistics [38–40], or are based on the Onsager theory of dissociation [35, 40]. For the present purpose of analysing the field ionization of deep impurities, it is sufficient to say that the proportionality, given by equation (69), is valid over a wide range of electric fields for both the classical model of Frenkel and the more sophisticated models referenced above.

If tunnelling occurs at higher fields, the role of a charge is reduced to increase the barrier transparency, because of the lowering of the barrier height. This gives only a correction to the tunnel ionization probability. In the limit of $\varepsilon_m > \mathrm{Ryd}^\star$ this correction has been calculated in [41], yielding a multiplicative factor for the emission rate $e(E)$ of equation (37):

$$e(E) = e_0 \exp\left[\frac{2\sqrt{2m^*\,\mathrm{Ryd}^\star}}{eE\tau_2}\ln\left(\frac{4\tau_2^3 e^2 E^2}{m^*\hbar}\right)\right]\exp\left(\frac{\tau_2^3 e^2 E^2}{3m^*\hbar}\right). \tag{70}$$

We readily see that the correction due to the impurity charge in equation (70) tends to unity with increasing electric field and becomes insignificant in strong fields. Thus, taking into account the Poole–Frenkel effect and phonon-assisted tunnelling ionization with the charge correction given in the above equation, we get that the logarithm of $e(E)$ varies as a function of the electric field at first like $\sqrt{E}$ and then changes to showing an $E^2$-dependence for high fields. The correction due to the charge in equation (70) approaches unity for increasing $E$ and therefore the charge correction becomes unimportant at high fields.

**Table 1.** Characteristics of the FIR laser lines used in this work.

| Wavelength ($\mu$m) | Frequency $\omega$ ($10^{12}$ s$^{-1}$) | Quantum energy $\hbar\omega$ (meV) | Line of CO$_2$ pump laser | Maximum intensity (kW cm$^{-2}$) | Medium |
|---|---|---|---|---|---|
| 76 | 25 | 16 | 10P(26) | 4000 | NH$_3$ |
| 90.5 | 21 | 14 | 9R(16) | 5000 | NH$_3$ |
| 148 | 13 | 8.5 | 9P(36) | 4500 | NH$_3$ |
| 250 | 7.5 | 4.9 | 9R(26) | 400 | CH$_3$F |
| 280 | 6.7 | 4.4 | 10R(8) | 1000 | NH$_3$ |
| 385 | 4.9 | 3.2 | 9R(22) | 5 | D$_2$O |
| 496 | 3.8 | 2.5 | 9R(20) | 10 | CH$_3$F |

## 4. Experimental methods

### 4.1. Optically pumped FIR molecular lasers

As a source of terahertz electric fields, a high-power pulsed FIR molecular laser pumped by a TEA CO$_2$ laser has been used [42]. Strong single-line emission has been achieved in the wavelength range from 30 to 500 $\mu$m applying NH$_3$, CH$_3$F, and D$_2$O as laser-active media. In table 1 the characteristics of these lines are listed together with lines of the TEA CO$_2$ laser which are used for pumping.

The photon energies corresponding to the wavelengths in the FIR lie in the 35–2 meV range and are in all cases substantially lower than the binding energies of the deep impurities in typical wide-gap semiconductors. The radiation pulse length varies for different lines from 10 to 100 ns. The radiated power was up to $\simeq$50 kW in the frequency range from $5 \times 10^{12}$ to $50 \times 10^{12}$ s$^{-1}$. The radiation could be focused to a spot of about 1 mm$^2$, with the maximum intensity reaching as high as 5 MW cm$^{-2}$ which corresponds to an electric field of about 40 kV cm$^{-1}$ inside the semiconductor samples.

The peak intensity of each single laser pulse could be monitored with fast noncooled photodetectors based on the photon drag effect [43], on intraband $\mu$-photoconductivity [44], or on the stimulated tunnelling effect in metal/semiconductor structures under plasma reflection [46]. The shape of the laser beam and the spatial distribution of the laser radiation were controlled with a Spiricon pyroelectric camera.

### 4.2. Samples

The tunnelling ionization processes were studied for two different types of deep impurity centre:

(i) substitutional impurities with weak electron–phonon coupling (Au, Hg, Cu, Zn in germanium, Au in silicon, and Te in gallium phosphide); and
(ii) off-site impurities with strong electron–phonon coupling where autolocalization occurs (Te in Al$_x$Ga$_{1-x}$As and in Al$_x$Ga$_{1-x}$Sb).

The thermal ionization energy of acceptor impurities $\varepsilon_T$ was in the case of germanium 150 meV (Au), 90 meV (Hg), 40 meV (Cu), and 30 meV (Zn), for Au in silicon, 300 meV, and for the donor tellurium in gallium phosphide, 90 meV [14]. Note that tellurium in gallium phosphide is basically a shallow impurity with a large central cell splitting of the hydrogen-atom-like ground state resulting in a large binding energy. These data and experimental results reviewed below are given in table 2. Doping with tellurium of Al$_x$Ga$_{1-x}$Sb with $x = 0.28$ and 0.5, and of Al$_x$Ga$_{1-x}$As with $x = 0.35$ yielded n-type conductivity. The samples showed all features characteristic of DX centres—in particular, persistent photoconductivity [16, 17, 47, 48].

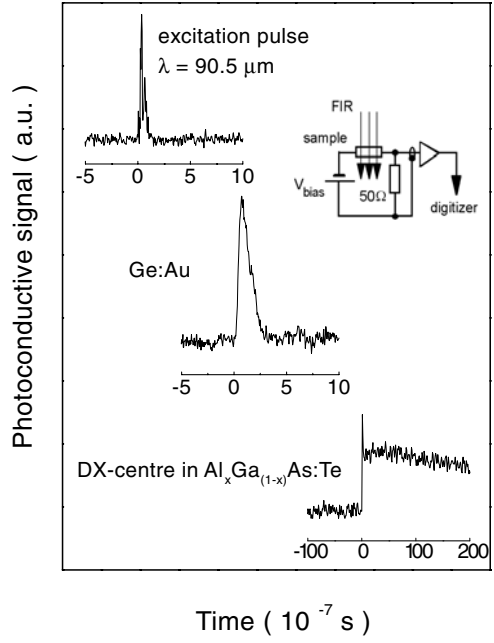

**Figure 10.** Oscillographic traces of the excitation pulse (top) at $\lambda = 90.5$ $\mu$m and of the photoconductive signals of Ge:Au (middle) and Al$_x$Ga$_{(1-x)}$As:Te (bottom, showing persistent photoconductivity at terahertz excitation!) The inset shows the measurement circuit.

**Table 2.** Parameters of samples investigated.

| | $\varepsilon_T$ (meV) | $\varepsilon_{opt}$ (meV) | $\Delta\varepsilon$ (meV) | $\tau_1$ ($10^{-15}$ s) | $\omega_{vib}$ ($10^{13}$ s$^{-1}$) | $S_{HR}$ |
|---|---|---|---|---|---|---|
| AlGaAs:Te | 140[a] | 850[a] | 710 | 3.3 | 25 | 4 |
| AlGaSb:Te | 120[a] | 860[a] | 740 | 29 | 3.0 | 36 |
| Ge:Au | 150[b] | 160 | 10 | 45 | 3.0 | 0.5 |
| Ge:Hg | 90[b] | 106 | 16 | 29 | 3.0 | 0.8 |
| Ge:Cu | 40[b] | — | — | 41 | — | — |

[a] Reference [13].
[b] Reference [14].

The resistance response of the semiconductor samples to laser irradiation was measured in a standard photoconductivity measurement circuit with a load resistance of $R_L = 50$ $\Omega$ (see the inset to figure 10). The bias voltage across the sample, $\leqslant 5$ V cm$^{-1}$, was substantially lower than the impurity avalanche breakdown threshold. The samples were placed in a temperature-controlled optical cryostat. Irradiation of the samples by visible and medium-infrared light was prevented by the use of crystalline quartz and black polyethylene filters, respectively. The measurements were carried out in the temperature range between 4.2 and 150 K, depending on the material, where at thermal equilibrium practically all carriers are bound to the impurities.

The ratio of the conductivity under illumination, $\sigma_i$, and the dark conductivity, $\sigma_d$, has been determined from peak values of photoconductive signals. For laser pulses shorter than the carrier capture time, as is the case here, $\sigma_i/\sigma_d$ is equal to $e(E)/e(0)$, where $e(E)$ is the emission rate as a function of the electric field strength $E$. Note that the terahertz response in the case of DX centres corresponds to the detachment of electrons from the defect, yielding persistent photoconductivity (see for details [8]).

## 5. Experimental results and discussion

### 5.1. Ionization of deep impurities by terahertz radiation

Semiconductors doped by deep impurities have been successfully used for a long time as low-temperature detectors for infrared radiation [49]. The long-wavelength limit to their use is determined by the binding energy of the impurities. At low irradiation intensities, no response is obtained from deep centres such as Ge:Au and Ge:Hg in the terahertz spectral range. However, on applying high-power FIR laser pulses to semiconductor samples doped with deep impurities, a photoconductive signal was observed, despite the fact that the pump photon energy was tens of times lower than the thermal ionization energy $\varepsilon_T$ [2–6, 8]. A signal rising superlinearly with the incident radiation intensity was found with all samples studied within the broad range of temperatures and wavelengths employed. The sign of the photoconductive signal corresponds to a decrease of the sample resistance. The characteristic decay times of the signal are different for different kinds of impurity and for different temperatures. The length of the photoresponse pulse for deep substitutional impurities is somewhat longer than that of the laser pulse (figure 10) and varies, depending on temperature, from 100 ns to 10 $\mu$s. These time constants correspond to the lifetimes of photoexcited carriers [14, 50, 51].

In the case of autolocalized DX centres in $Al_xGa_{1-x}As$ and $Al_xGa_{1-x}Sb$, an increase of the sample conductivity is observed which persists for several hundreds of seconds after the excitation pulse (figure 10). This behaviour is characteristic of the decay of persistent photoconductivity in semiconductors with DX centres [13].

The increase of the sample conductivity due to high-power terahertz radiation pulses can be either due to radiation absorption by free carriers (electron gas heating, $\mu$-photoconductivity [52]) or, on the other hand, due to the generation of additional free carriers by impurity ionization. We discuss first the possible effect of heating of the lattice or of the electron gas, since this is the most natural mechanism of photoconductivity due to intense illumination. Carrier heating was studied in detail in the submillimetre range on samples with shallow impurities and at not-too-low temperatures, i.e. where the impurities are ionized and the conditions are most favourable for heating. As a result of this investigation, in the case of excitation of samples with deep impurities, the possibility of electron gas heating as the cause of the observed impurity ionization can be excluded [5]. This conclusion is mainly based on the kinetics of the detected signals. The photoresponse signal due to electron heating should either reproduce the shape of the pump pulse or be more complex but not longer than the pump pulse [5, 53, 54]. In contrast, the observed signals are substantially longer than the pump pulses and their durations agree with the recombination times of excited free carriers. The detachment of electrons in samples containing DX centres is also very clearly demonstrated by the FIR photoconductivity which persists when the electrons are definitely at ambient temperature. Furthermore, electron gas heating in the temperature range and free carrier concentration range investigated here should produce negative photoconduction, whereas the photoconductivity observed experimentally is positive. Thus, the observation of positive photoconductive signals with decay times substantially in excess of the pump pulse length excludes the possibility of electron gas heating and the corresponding photoconductivity being the mechanism of the observed photoresponse. Therefore the photoresponse must indeed be caused by photoionization of deep impurities by irradiation with the photon energy $\hbar\omega$ much less than the thermal ionization energy of impurities $\varepsilon_T$.

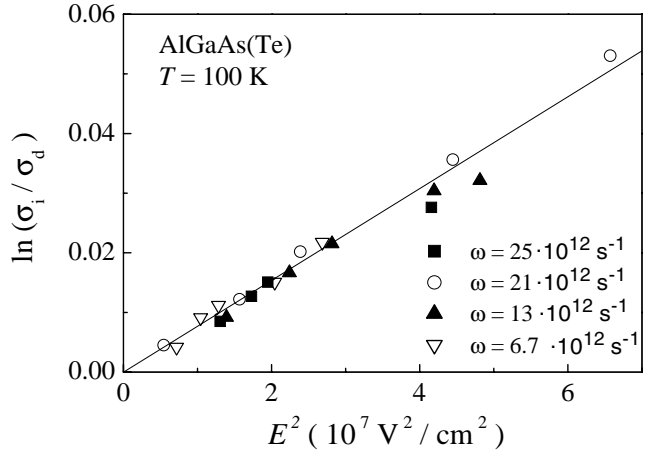

**Figure 11.** The logarithm of the ratio of the irradiated conductivity, $\sigma_i$, to the dark conductivity, $\sigma_d$, as a function of the squared electric field, $E^2$ for $Al_{0.35}Ga_{0.65}As$:Te at $T = 100$ K for various frequencies $\omega$.

Figure 11 displays the dependence of the photoresponse of the DX centre in AlGaAs on the square of the radiation electric field strength for four different wavelengths. The experimentally determined relative change in conductivity, $\Delta\sigma/\sigma_d = (\sigma_i - \sigma_d)/\sigma_d$, corresponds to the relative change in the free carrier concentration, which, in turn, is proportional to the change in the impurity ionization probability.

Detailed investigations of the photoconductivity with high-power terahertz excitation of semiconductors doped with deep impurities led to the conclusion that the carrier detachment process is caused by tunnelling [2, 3, 8]. The measurements showed that at relatively high temperatures, photoconductivity does not depend on radiation wavelengths at long $\lambda$ throughout the intensity range covered by the present investigation. This is demonstrated by figure 11 which shows that the curves for all wavelengths coincide within the measurement accuracy.

An increase in frequency or decrease in temperature result in the onset of a frequency dependence of the ionization probability. As shown in section 3.2, the limit between frequency-independent and frequency-dependent ionization of deep impurities is given by the tunnelling time $\tau_2$. The ionization probability is independent of frequency in the quasi-static limit as long as $\omega\tau_2 \leqslant 1$. In the high-frequency limit, $\omega\tau_2 > 1$, the ionization probability drastically increases. This increase indicates the transition from semiclassical physics, where ionization is accomplished by the classical electric field amplitude, to the full quantum mechanical process of multiphoton transitions.

### 5.2. Phonon-assisted tunnelling ionization

*5.2.1. The quasi-static limit.* Phonon-assisted tunnelling ionization is characterized by an exponential dependence on the squared-wave electric field amplitude: $e(E) = e(0)\exp(E^2/E_c^2)$ (see equation (58)). Such an increase in the photoconductive signal has been observed for all samples within a broad range of fields and temperatures, the ranges being different for different samples and frequencies. The experimental dependences of $\ln(\sigma_i/\sigma_d)$ on the squared amplitude of the electric field are shown in figure 11 for $Al_xGa_{1-x}As$ and for Ge:Au in figure 12. The measurements were performed at different temperatures and wavelengths. We see that for each temperature there exists a field interval within which the probability of photoionization depends on the electric field amplitude as $\exp(E^2/E_c^2)$. A comparison of experimental data on terahertz ionization of the Au impurity in Si at $T = 300$ K with earlier studies of the dependence of thermal ionization probability on a dc electric field, $e(E)$, made by means of capacitive spectroscopy [55, 56] showed that in both cases $e(E) \propto \exp(E^2/E_c^2)$, with the values of $E_c$ differing by a factor of 1.5–2. This may be considered a good agreement between the results obtained by such different methods, if we take into account the field inhomogeneities present in a sample studied by DLTS.

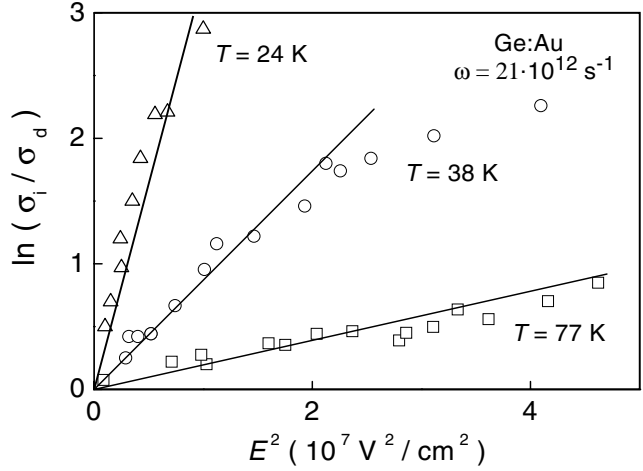

**Figure 12.** $\ln(\sigma_i/\sigma_d)$ for Ge:Au ($\varepsilon_T = 150$ meV) versus $E^2$ at $\lambda = 90.5$ $\mu$m for various temperatures $T$. The straight lines are fitted to $e(E) \propto \exp(E^2/E_c^2)$ with $E_c^2$ as a fitting parameter.

Figures 11 and 12 show also plots of the $A\exp(E^2/E_c^2)$ relation calculated with the fitting parameter $E_c^2$. As follows from equations (58) and (59), the slope of the experimental curves in the field region where $\ln(\sigma_i/\sigma_d) \propto \exp(E^2/E_c^2)$ permits determination of the tunnelling times $\tau_2$. In order to extract $\tau_2$ from experimental data, one has to know the effective carrier mass, which determines the tunnelling process. In figure 13, the tunnelling time $\tau_2$ is shown as a function of reciprocal temperature for a number of deep impurities studied. In the case of deep acceptors in germanium, the light-hole mass was used. Figure 13 demonstrates the good agreement of the experimental values of $\tau_2$ with equation (63). One may thus conclude that holes bound to a deep acceptor tunnel into the light-hole subband [3]. This is due to the fact that the symmetry of substitutional impurities corresponds to the point group $T_d$, and that the ground state of a deep impurity represents a superposition of the light- and heavy-hole states. Thus an acceptor-bound hole can be associated with neither the light nor the heavy mass. It was shown theoretically [57] that tunnelling depends essentially on the wavefunction tail distant from the centre, and that it is the light holes that provide a major contribution to this tail.

For comparison, figure 13 also shows a plot of $\hbar/2k_BT$. We see that $\tau_2$ is of the order of $\hbar/2k_BT$. Note, however, an essential point. As is evident from the experimental data presented in figure 13, for any temperature, $\tau_2$ is larger than $\hbar/2k_BT$ for substitutional impurities, but less than $\hbar/2k_BT$ for autolocalized DX centres. This result is in excellent agreement with theory (see equation (63)). Thus, by determining the tunnelling time from data on phonon-assisted tunnelling ionization in a high-frequency electric field but in the quasi-static limit, one can unambiguously identify the type of deep-impurity adiabatic potential [8]. The temperature-independent tunnelling times $\tau_1 = \tau_2 - \hbar/2k_BT$ are given in table 2 for different impurities.

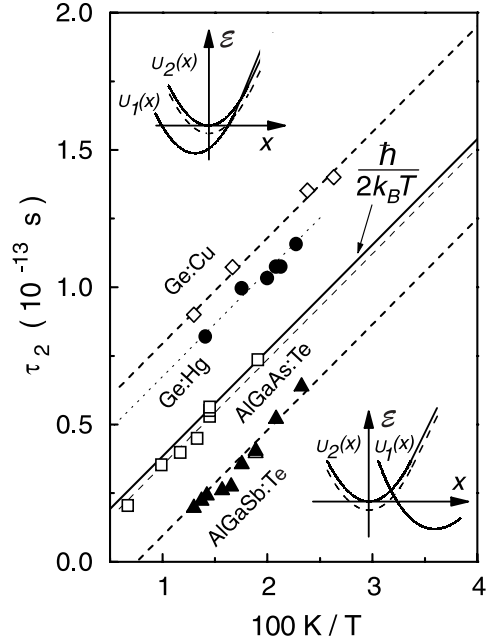

**Figure 13.** Tunnelling time $\tau_2$ derived from experimental values of $E_c^2$ versus reciprocal temperature for substitutional impurities (Ge:Cu and Ge:Hg) and DX centres (AlGaAs:Te and AlGaSb:Te). The full line represents $\hbar/2k_BT$ and the dashed lines are fits to $\tau_2 = \hbar/2k_BT \pm \tau_1$. The line $\hbar/2k_BT$ separates the range of weak electron–phonon interaction ($\tau_2 = \hbar/2k_BT + \tau_1$) and strong electron–phonon interaction ($\tau_2 = \hbar/2k_BT - \tau_1$). The corresponding adiabatic potentials are shown in the insets (top left and bottom right). The tunnelling time $\tau_1$ is of the order of $10^{-14}$ s (see table 2).

Concluding, we note that the specific structure of the adiabatic potentials of DX centres allows a process inverse to phonon-assisted tunnelling detachment of carriers in terahertz fields. Irradiating samples with visible radiation accumulates free carriers in the conduction band with the bottom $U_2(x)$. At low temperatures ($T < 100$ K) the lifetime of these carriers is very long, and is responsible for the persistent photoconductivity. Illuminating the samples in this state with terahertz pulses produces, in contrast to the positive photoconductive signal without pre-illumination, a negative photoconductive signal caused by phonon-assisted tunnelling from $U_{2\varepsilon}(x)$ to $U_1(x)$ and subsequent capture in the impurity bound state by phonon emission.

*5.2.2. The high-frequency limit.* The frequency-independent tunnelling is limited to frequencies $\omega$ with $\omega\tau_2 < 1$ (see equation (60)). The fact that the tunnelling time $\tau_2$ depends on temperature (see figure 13) allows one to proceed into the high-frequency regime $\omega\tau_2 > 1$ simply by cooling the samples. This is an important advantage in this case, because the other opportunity to get $\omega\tau_2 \geqslant 1$, i.e. raising the frequency, is limited by one-photon absorption. The measurements show that in a finite electric field range for the case of $\omega\tau_2 \geqslant 1$, the ionization probability still depends exponentially on the square of the electric field strength $e(E) \propto \exp(E^2/E_c^{*2})$. The essential difference compared to the $\omega\tau_2 < 1$ situation is that the characteristic field $E_c^*$ now becomes frequency dependent. It is found that ionization is enhanced with rising frequency. This behaviour has been observed for all impurities and is demonstrated for AlGaAs:Te in figure 14 and for Ge:Hg in figure 15 for temperatures that are not very low and electric field strengths that are not very high. At higher field strengths the exponential dependence on $E^2$ ceases and the ionization probability rises more slowly with increasing $E$. This high-field case will be discussed below.

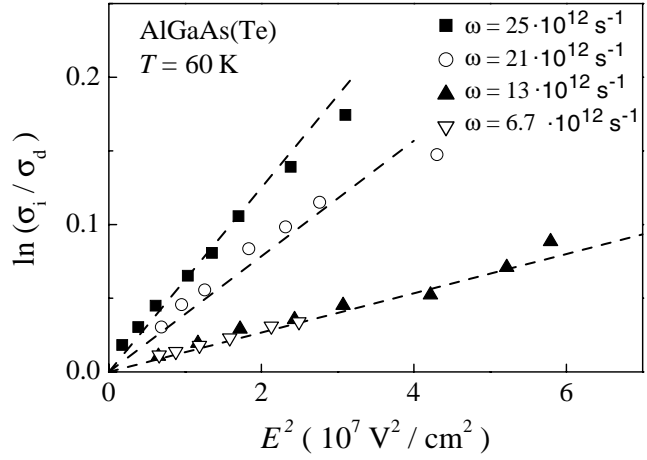

**Figure 14.** $\ln(\sigma_i/\sigma_d)$ for $Al_{0.35}Ga_{0.65}As$:Te as a function of $E^2$ for different frequencies $\omega$ at $T = 60$ K.

The experimentally determined values of $E_c^*$ for various frequencies, temperatures, and materials allow one to obtain the value of $\tau_2^*/\tau_2$. Figure 5 shows this ratio as a function of $\omega\tau_2$ in comparison to calculations after equation (60). The tunnelling times $\tau_2$ were determined from frequency-independent values of $E_c^* = E_c$. The experimental results shown in figure 5 are grouped according to the materials. For each material the variation of the value of $\omega\tau_2$ has been obtained by applying different radiation frequencies in the range from $3.4 \times 10^{12}$ to $25 \times 10^{12}$ s$^{-1}$ and for different temperatures between 4.2 and 150 K. A good agreement between theory and experiment is obtained. It should be pointed out that the theory does not contain any fitting parameter.

The enhancement of tunnelling at frequencies higher than the inverse tunnelling time has been anticipated in a number of theoretical works [1, 58–60], but has been demonstrated experimentally only recently [2]. In contrast to the case for static electric fields where the electron tunnels at a fixed energy, in alternating fields the energy of the electron is not conserved during tunnelling. In this case the electron can absorb energy from the field (see the inset in figure 5) and hence leaves the impurity at an effectively narrower tunnelling barrier. This leads to a sharp increase of the tunnelling probability with increasing frequency. The observed enhancement of the ionization probability demonstrates that an electron can indeed absorb energy below a potential barrier if the process of tunnelling is induced by a high-frequency alternating electric field. The absorption of energy is controlled by the electron tunnelling time $\tau_e$, i.e. the Büttiker–Landauer time. In the case of phonon-assisted tunnelling, the energy of the electron under the barrier, $\varepsilon_m$, follows from the condition that the electron tunnelling time $\tau_e$ is equal to the defect tunnelling time $\tau_2$, which is determined by the tunnelling reconstruction of the defect vibration system equation (63). Thus the Büttiker–Landauer time of electron tunnelling can be varied by the temperature and can be measured from the field dependence of the ionization probability.

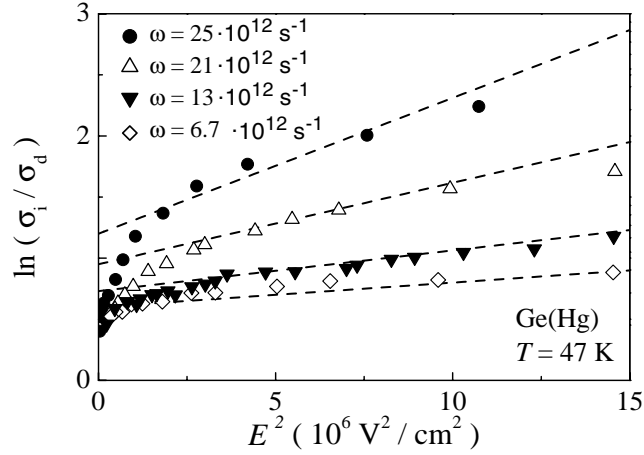

**Figure 15.** $\ln(\sigma_i/\sigma_d)$ for Ge:Hg as a function of $E^2$ for different frequencies at $T = 47$ K.

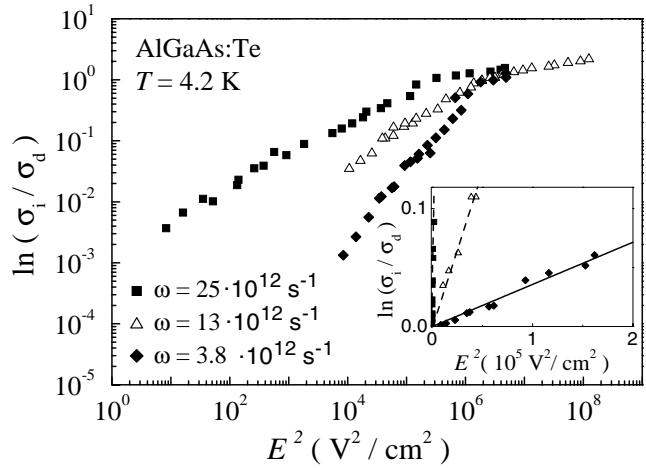

**Figure 16.** $\ln(\sigma_i/\sigma_d)$ for $Al_{0.35}Ga_{0.65}$As:Te as a function of $E^2$ for different frequencies $\omega$ at liquid helium temperature. The inset shows the low-field behaviour.

Further decrease of the temperature increases the tunnelling time and leads to a much stronger frequency dependence of the ionization probability. Figure 16 shows measurements carried out at 4.2 K on AlGaAs:Te. The data have been obtained making use of persistent photoconductivity of the sample in order to reduce the dark resistance [61]. Because of the large binding energy of the DX centre, the resistance of the sample cooled in the dark is too high for detecting any signal at this temperature in response to terahertz radiation. Therefore the sample was illuminated for a short time with weak visible and near-infrared light detaching the electrons from a small fraction of DX centres. As a result, the sample resistance drops to experimentally reasonable values of the order of several hundred megohms. After this

27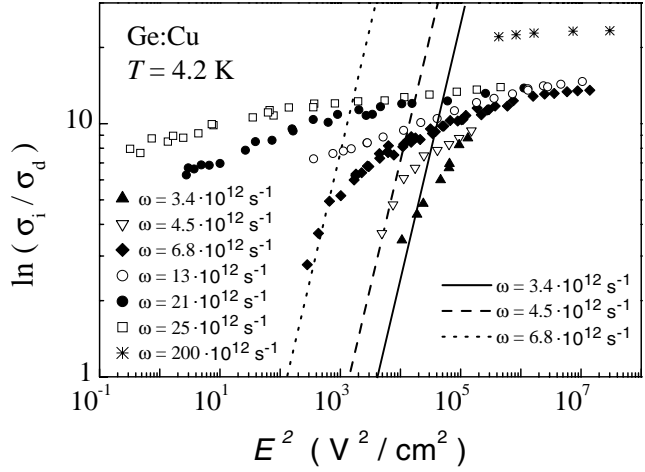

**Figure 17.** $\ln(\sigma_i/\sigma_d)$ for Ge:Cu as a function of $E^2$ for different frequencies $\omega$ at liquid helium temperature. Lines show calculations after equations (58)–(60) for the three lowest frequencies used in experiments.

illumination the resistance change at 4.2 K is persistent and allows one to measure the process of electron detachment by terahertz radiation.

In order to display in one figure the total set of data covering eight orders of magnitude in the square of the electric field strength, $\log(E^2)$ has been plotted on the abscissa. To make an easy comparison to the $\exp(E^2/E_c^2)$ dependence of $\sigma_i/\sigma_d$ possible, a log–log presentation has been used for the ordinate. In the low-field range the characteristic $\propto \exp(E^2/E_c^{*2})$ field dependence of phonon-assisted tunnelling is observed. This is additionally shown in the inset of figure 16 in a log–linear plot. The frequency dependence in the field range of phonon-assisted tunnelling is so strong that a change of three orders of magnitude of $E^2$ needs only a six times change in the frequency $\omega$.

Similar results have been obtained for substitutional impurities having a smaller binding energy and showing larger tunnelling times $\tau_2$ (see figure 13). Figure 17 shows experimental results for Ge:Cu at $T = 4.2$ K in the frequency range between 3.4 and 25 THz. Here the frequency dependence at low field strengths is even stronger. For a given constant signal a change of six orders of magnitude of $E^2$ needs a factor-of-seven change in frequency $\omega$.

*5.2.3. The transition to direct tunnelling.* At higher field strengths the field dependence of the emission probability gets much weaker and the frequency dependence practically disappears. The transition to frequency-independent probability at higher field strength occurs at lower fields for Ge:Cu than for DX centres in AlGaAs:Te. The weak increase of the frequency-independent carrier emission at high electric fields cannot be attributed to emptying of the impurity states. This has been proved by one-photon ionization of Ge:Cu using $CO_2$ laser radiation of $\omega = 200 \times 10^{12}$ s$^{-1}$. The saturation level of the photoconductivity where practically all impurities are ionized lies well above the terahertz data (see figure 17).

This complex dependence of the ionization probability on the field strength and radiation frequency is a result of the transition from phonon-assisted tunnelling at low field strengths to direct tunnelling without phonons at high fields (see section 3.3 and figure 7). At low field strength the electric field and frequency dependence are controlled by $\tau_2$ being independent of the electron energy $\varepsilon$. At high fields, these tunnellings become dependent on the electron energy and will therefore be denoted by $\tau_{2\varepsilon}(\mathcal{E}, \varepsilon)$ and $\tau_e(\varepsilon)$. The emission probability for phonon-assisted tunnelling as a function of the electric field strength given by equation (3) was obtained in the limit where corrections to the thermal emission resulting from the electron tunnelling are small, i.e., the energy of electron tunnelling $|\varepsilon_m|$ is much smaller than the defect tunnelling energy $\mathcal{E}_m$. In the opposite limit, $|\varepsilon_m| \geqslant \mathcal{E}_m$, direct carrier tunnelling from the ground state into the continuum, without participation of phonons, becomes dominant [4–6, 62]. Direct electron tunnelling occurs at the crossing of the $U_{2\varepsilon}(x)$ and $U_1(x)$ potential curves, where an electronic transition is possible without any change in the configuration coordinate. This effect, leading to weaker field dependence of the ionization probability in comparison to that of phonon-assisted tunnelling, dominates the ionization process at very high fields.

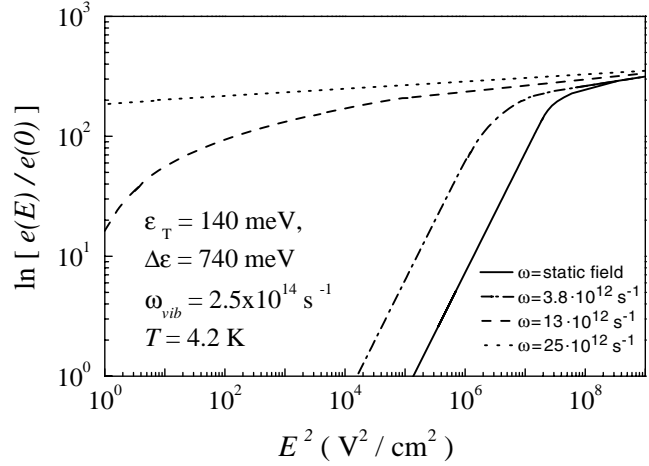

**Figure 18.** The logarithm of the normalized ionization probability versus $E^2$ calculated for different frequencies used in experiments. Calculations were carried out for 4.2 K using the parameters of $Al_{0.35}Ga_{0.65}As$:Te taking into account phonon-assisted tunnelling and direct tunnelling.

The results of calculations of the ionization probability over a wide range of electric field strength, which demonstrate the transition from phonon-assisted tunnelling to direct tunnelling, are presented at figures 18. The calculations are performed for the Huang–Rhys adiabatic potential model (see equations (3) and (4)). The probability of tunnelling ionization has been calculated by using equation (35). For the calculations, the defect tunnelling times, the electron tunnelling time, and the values of the optimal defect and electron tunnelling energies are needed. The defect tunnelling times $\tau_{2\varepsilon}(\mathcal{E}, \varepsilon)$ and $\tau_{1\varepsilon}(\mathcal{E}, \varepsilon)$ as a function of the electron energy $\varepsilon$ and the defect energy $\mathcal{E}$ have been calculated after equations (3), (4), and (37). The electron tunnelling time as a function of electron energy $\varepsilon$, electric field strength $E$, and radiation frequency $\omega$ has been obtained using equation (56) for $E_2 = 0$ and $E_1 = E$ which corresponds to linearly polarized radiation. Optimal electron and defect tunnelling energies, $\varepsilon_m$ and $\mathcal{E}_m$, have been obtained using equations (36) and (38). Figure 19 shows the results of the calculations using the parameters of AlGaAs:Te and Ge:Cu (ignoring Coulomb interaction). The calculations, which take into account both processes (phonon-assisted tunnelling and direct tunnelling), were carried out for several field frequencies used in the experiments.

The theory qualitatively describes the whole of the complex features of the tunnelling ionization probability as a function of frequency and electric field strength. The experimentally observed stronger frequency dependence at low field strength of the ionization probability of Ge:Cu compared to that of AlGaAs:Te is caused by the larger values of $\tau_2$ in the first case.

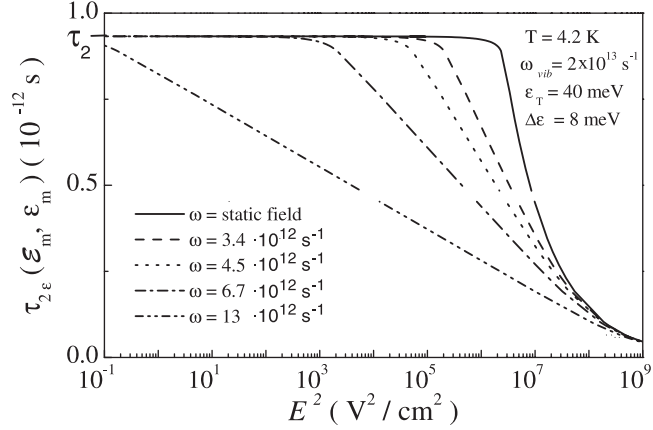

**Figure 19.** The tunnelling time $\tau_{2\varepsilon}(\mathcal{E}_m, \varepsilon_m)$ versus $E^2$ calculated for different frequencies used in experiments. Calculations were carried out for 4.2 K using the parameters of Ge:Cu taking into account phonon-assisted tunnelling and direct tunnelling but ignoring Coulomb interaction.

The disappearance of the frequency dependence at very high fields is caused by the reduction of tunnelling time $\tau_{2\varepsilon}(\mathcal{E}_m, \varepsilon_m)$ with the rising electric field strength. In figure 19 the electric field dependence of $\tau_{2\varepsilon}(\mathcal{E}_m, \varepsilon_m)$ calculated with the parameters of Ge:Cu is shown. Figure 19 presents data for 4.2 K and for various frequencies. The physical reason for the drop of $\tau_{2\varepsilon}(\mathcal{E}_m, \varepsilon_m)$ is the increase of the electron tunnelling energy $\varepsilon$. Because of this, $\omega\tau_{2\varepsilon}(\mathcal{E}_m, \varepsilon_m)$ becomes smaller than one and thus the frequency dependence vanishes.

The feature of the exponential dependence on the square of the electric field strength changing to a weaker field dependence at lower fields for Ge:Cu compared to AlGaAs:Te is caused by the difference of the binding energies. The transition from phonon-assisted tunnelling to direct tunnelling depends substantially on the value of the binding energy (equation (66)). For smaller binding energies it occurs at lower fields, yielding weaker field and frequency dependences.

Note that the results of calculations obtained in the framework of the Huang–Rhys model cannot be used for a quantitative description of the transition from the phonon-assisted to the direct tunnelling regime because the real shape of the potentials can differ from the parabolic shape used in the model of Huang and Rhys. Furthermore, to achieve a quantitative agreement of theory and experiment, heating of the phonon system by energy transfer from the electrons should be taken into account. The efficient tunnelling ionization at high fields causes a substantial increase in free carrier concentration. Thus, free electrons may be heated by terahertz radiation due to Drude absorption. An increase of the sample temperature of just a few degrees leads to a decrease of $\tau_{2\varepsilon}$. As a result, the normalized emission rate $e(E)/e(0)$ decreases and the frequency dependence of the emission probability becomes much weaker. Furthermore, for the case of charged impurities one needs to extend the theory by taking into account the lowering of the barrier height in the presence of an external electric field due to the Coulomb field of impurities.

### 5.3. The Poole–Frenkel effect in high-frequency fields

In the region of relatively weak electric fields, one also observes deviations from the $\exp(E^2/E_c^2)$ behaviour of the phonon-assisted tunnelling [19, 63], which is clearly seen from figure 20 displaying the $\ln(\sigma_i/\sigma_d)$ versus $E^2$ relation for Ge:Hg. The dominant mechanism in this ionization process is the Poole–Frenkel effect (see section 3.4). Data for the weak-field region are shown in the inset of figure 20, where $\ln(\sigma_i/\sigma_d)$ is plotted as a function of the square root of the high-frequency electric field amplitude, $\sqrt{E}$. In the low-field range, the ionization probability is seen to follow closely the $e(E) \propto \exp\sqrt{E/E_{PF}}$ relation. The square-root dependence of $\ln(\sigma_i/\sigma_d)$ on $E$ and its temperature behaviour are in good agreement with equations (68) and (69) describing the Poole–Frenkel effect.

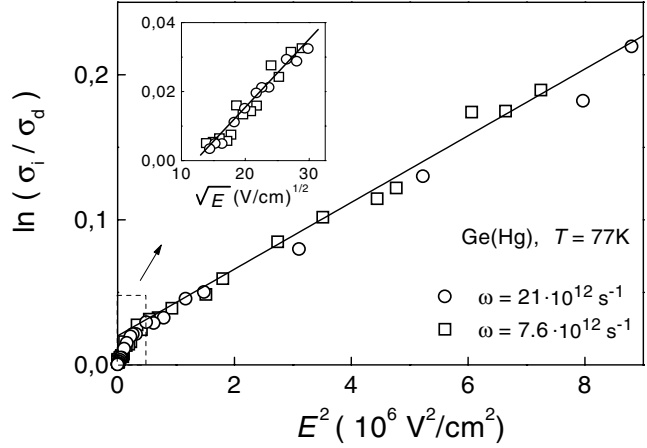

**Figure 20.** $\ln(\sigma_i/\sigma_d)$ for Ge:Hg as a function of $E^2$ for different frequencies $\omega$ at $T = 47$ K.

The charge effect manifests itself also in the phonon-assisted tunnelling ionization, resulting in an additional factor in the ionization probability according to equation (70). This is seen from extrapolation of the straight lines corresponding to the region of phonon-assisted tunnelling ionization to zero electric field. We see that $\ln(\sigma_i/\sigma_d)$ does not vanish for $E = 0$ (figures 15 and 20), which implies that $\sigma_i$ is not equal to $\sigma_d$, as this follows from equation (58) which does not take into account the charge effect.

Finally we note that the Poole–Frenkel effect has not been observed with DX centres within the accuracy of the experiment. The emission probability can consistently be described as phonon-assisted tunnelling down to vanishing electric field strength. As seen from figures 11 and 14, the logarithm of the normalized emission probability, $\ln(e(E)/e(0)) = \ln(\sigma_i/\sigma_d)$, can be extrapolated to zero at zero electric field $E$ with constant slope. This observation proves that there is no Coulomb force between the detached electron and the impurity centre. Thus, the DX ground state is negatively charged while the electron-detached state is neutral.

The determination of the slope of the power law of $\ln(e(E))$ versus $E$ for small fields is an easy and unambiguous way to determine the nature of the field enhancement of the carrier emission [19]. If phonon-assisted tunnelling prevails even for low fields, the carrier is emitted from a neutral impurity. Since an enhancement of the carrier emission rate in an electric field is observed for both charged and neutral impurities, the observation of the enhancement of the emission rate in an electric field alone is not sufficient for reaching a conclusion on the charge state of an impurity. The latter can be inferred from a plot of the logarithm of the emission probability $e(E)$ as a function of $E^2$ at high fields and $\sqrt{E}$ at low field strength or from plotting the logarithm of the normalized emission probability $e(E)/e(0)$ versus $E^2$ in the electric field range of phonon-assisted tunnelling.

## 6. Summary

In summary, tunnelling ionization of deep impurities in semiconductors by high-intensity submillimetre laser radiation with photon energies much smaller than the impurity ionization energy has been investigated in theory and experiment. The measurements have been carried out over a broad range of intensities, wavelengths, and temperatures, and for a variety of impurities.

It has been shown that in a wide range of experimental conditions the electric field laser radiation in the terahertz frequency range acts like a dc field. The effect of an electric field on the thermal emission and capture of carriers is of importance for the kinetics and dynamics of semiconductors. A high static electric field drives the system into avalanche breakdown which is usually associated with a large increase in noise, self-generated oscillations, and current filamentations. These effects substantially change the properties of the material and disguise the elementary properties of tunnelling. The present method of ionizing impurities by short laser pulses avoids these problems. The use of short high-power terahertz laser pulses permits contactless application of very high electric field strengths. The radiation pulse is shorter than the time needed to form a free carrier avalanche; therefore extremely high electric field strengths may be applied. The dc bias field required to record photoconductivity may be kept well below the threshold of instability where the perturbation of the electron system is small, avoiding injection at the contacts. The intrinsically high sensitivity of photoconductivity gives a measurable signal from a few radiation-excited carriers.

In contrast to tunnelling ionization of atoms by applying very short high-power pulses of visible lasers, tunnelling ionization of atom-like centres in solids strongly depends on the electron–phonon interaction with the thermal bath over a wide range of electric field strength. Only at very high electric fields does tunnelling occur, like in the case of atoms, directly, without involving phonons. With charged impurities at small electric field strengths, the Poole–Frenkel effect is superimposed on the tunnelling ionization process. Therefore three characteristic electric field dependences of the impurity ionization may be distinguished:

 (i) at low fields, the Poole–Frenkel emission due to lowering of the thermal ionization energy of attractive impurity centres;
 (ii) at higher field, phonon-assisted tunnelling; and
(iii) at very high fields, direct tunnelling.

In the range of phonon-assisted tunnelling, the electron–phonon interaction determines the time of tunnelling due to the reconstruction of the vibrational system by the tunnelling detachment of a carrier. Hence the tunnelling time can be easily varied by changing the temperature. The field dependence of the photoconductive signal allows one to determine defect tunnelling times, the Huang–Rhys parameter, the structure of the adiabatic potentials, and the defect charge. Thus, tunnelling ionization may be applied to characterize deep impurities, complementing the usual method of DLTS.

The observed enhancement of the tunnelling probability on increasing the frequency gives evidence that the tunnelling carrier can absorb energy while tunnelling under the barrier. The enhancement of the tunnelling occurs when the frequency becomes larger than the inverse tunnelling time. This limit is approached in the terahertz regime. Applying a magnetic field normal to the electric field vector, increase of temperature and/or increase of electric field lead to a shift of the boundary between the quasi-static and frequency-dependent regimes to higher frequencies.

## Acknowledgments

We thank V I Perel for many helpful discussions. Financial support by the Deutsche Forschungsgemeinschaft and the Russian Foundation of Fundamental Investigations are gratefully acknowledged.